\documentclass[prb,superscriptaddress,twocolumn,showpacs,dvips]{revtex4}
\usepackage{feynmp}
\usepackage{amssymb}
\usepackage{epsfig}
\usepackage{graphicx}
\usepackage{subfigure}
\usepackage{hyperref}
\usepackage{bbm}
\usepackage{gensymb}
\usepackage{float}
\usepackage{color}

\begin{document}

\title{Connection between in-plane upper critical field $H_{c2}$ and
gap symmetry in layered $d$-wave superconductors revisited}

\author{Jing-Rong Wang}
\altaffiliation{wangjr@mail.ustc.edu.cn} \affiliation{Department of
Modern Physics, University of Science and Technology of China,
Hefei, Anhui 230026, P. R. China} \affiliation{High Magnetic Field
Laboratory, Hefei Institutes of Physical Science, Chinese Academy of
Sciences, Hefei 230031, P. R. China}

\author{Guo-Zhu Liu}
\altaffiliation{gzliu@ustc.edu.cn} \affiliation{Department of Modern
Physics, University of Science and Technology of China, Hefei, Anhui
230026, P. R. China}

\author{Chang-Jin Zhang}
\affiliation{High Magnetic Field Laboratory, Hefei Institutes of
Physical Science, Chinese Academy of Sciences, Hefei 230031, P. R.
China} \affiliation{Collaborative Innovation Center of Advanced
Microstructures, Nanjing University, Nanjing 210093, P. R. China}

\begin{abstract}
Angle-resolved upper critical field $H_{c2}$ provides an efficient
tool to probe the gap symmetry of unconventional superconductors. We
revisit the behavior of in-plane $H_{c2}$ in $d$-wave
superconductors by considering both the orbital effect and Pauli
paramagnetic effect. After carrying out systematic analysis, we show
that the maxima of $H_{c2}$ could be along either nodal or antinodal
directions of a $d$-wave superconducting gap, depending on the
specific values of a number of tuning parameters. This behavior is
in contrast to the common belief that the maxima of in-plane
$H_{c2}$ are along the direction where the superconducting gap takes
its maximal value. Therefore, identifying the precise $d$-wave gap
symmetry through fitting experiments results of angle-resolved
$H_{c2}$ with model calculations at a fixed temperature, as widely
used in previous studies, is difficult and practically unreliable.
However, our extensive analysis of angle-resolved $H_{c2}$ show that
there is a critical temperature $T^{*}$: in-plane $H_{c2}$ exhibits
its maxima along nodal directions at $T < T^{*}$ and along antinodal
directions at $T^{*} < T < T_c$. The concrete value of $T^{*}$ may
change as other parameters vary, but the existence of $\pi/4$ shift
of $H_{c2}$ at $T^{\ast}$ appears to be a general feature. Thus a
better method to identify the precise $d$-wave gap symmetry is to
measure $H_{c2}$ at a number of different temperatures, and examine
whether there is a $\pi/4$ shift in its angular dependence at certain
$T^{*}$. We further show that Landau level mixing does not change
this general feature. However, in the presence of
Fulde-Ferrell-Larkin-Ovchinnikov state, the angular dependence of
$H_{c2}$ becomes quite complicated, which makes it more difficult to
determine the gap symmetry by measuring $H_{c2}$. Our results
indicate that some previous studies on the gap symmetry of
CeCu$_{2}$Si$_{2}$ are unreliable and need to be reexamined, and
also provide a candidate solution to an experimental discrepancy in
the angle-resolved $H_{c2}$ in CeCoIn$_{5}$.
\end{abstract}


\maketitle


\section{Introduction\label{Sec:Introduction}}

Identifying the precise gap symmetry is generically regarded as an
important step on the road of searching for the microscopic pairing
mechanism of unconventional superconductivity \cite{Norman11,
Scalapino12}. Different from the isotropic, phonon mediated BCS
superconductors, unconventional superconductors are believed to be
induced by the strong electron correlations and normally possess an
anisotropic non-$s$-wave superconducting gap. Extensive theoretic
studies have found that an anisotropic superconducting gap always
leads to an anisotropic, angle dependent in-plane upper critical
field $H_{c2}$ \cite{Gorkov84, Won94, Takanaka95}. Motivated by
these studies, the angle-resolved in-plane $H_{c2}$ has recently
been widely used to determine the gap symmetry of a number of
unconventional superconductors, including cuprate superconductors
\cite{Koike96, Naito01}, heavy fermion superconductors \cite{Won04,
Weickert06, Vieyra11}, iron based superconductors \cite{Murphy13},
and other types of superconductors such as Sr$_{2}$RuO$_{4}$
\cite{Mao00} and K$_{2}$Cr$_{3}$As$_{3}$ \cite{Zuo15}.

It is widely accepted that cuprate superconductors have a
$d_{x^2-y^2}$ wave gap \cite{Tsuei00, Damascelli03}. However, the
precise gap symmetry of many heavy fermion superconductors is still
unclear. Among the dozens of known heavy fermion compounds,
CeCu$_{2}$Si$_{2}$ and CeCoIn$_{5}$ have attracted special
experimental and theoretical interest \cite{Lohneysen07, Stockert12, Matsuda06,
Sarrao07, Thompson12}.

As the first heavy fermion superconductor \cite{Steglich79},
CeCu$_{2}$Si$_{2}$ has been studied for more than three decades, but
no consensus has been reached concerning its precise gap symmetry. A
number of earlier experiments provides evidence for a $d_{x^2 -
y^2}$-wave gap \cite{Stockert08, Eremin08}. Subsequent studies of
angle-resolved $H_{c2}$ by Vieyra \emph{et al.} \cite{Vieyra11}
found that the in-plane $H_{c2}$ exhibits a fourfold oscillation
with its maxima being along the $[100]$ direction. By fitting model
computations to their measurements, Vieyra \emph{et al.}
\cite{Vieyra11} proposed that the gap symmetry of CeCu$_{2}$Si$_{2}$
should be $d_{xy}$-wave, which is in sharp contrast to most previous
works \cite{Stockert08, Eremin08}. Recent specific heat measurements
suggested that CeCu$_{2}$Si$_{2}$ may have a nodeless multi-band
superconducting gap \cite{Kittaka14}, which challenges the widely
accepted notion that the gap symmetry of this superconductor is
$d$-wave. Observations made in the vortex state by scanning
tunneling microscopy and spectroscopy are consistent with a
multi-band gap with nodes \cite{Enayat15}. First-principle
calculations speculated that a promising pairing state might be
multi-band $s_{\pm}$-wave with loop shaped nodes \cite{Ikeda15}.
Moreover, by measuring the change of penetration depth and
renormalized superfluid density, and then comparing these findings
to previous measurements of specific heat, a nodeless $d+d$
band-mixing state was also proposed as a candidate for the gap
symmetry of CeCu$_{2}$Si$_{2}$ \cite{Pang16}.

CeCoIn$_{5}$, discovered in 2001 by Petrovic \emph{et al.}
\cite{Petrovic01}, is known to have one of the highest critical
temperature, roughly $T_{c} = 2.3$K, among the whole heavy fermion
family. Many experimental measurements, including thermal
conductivity \cite{Izawa01}, specific heat in rotated magnetic field
\cite{An10}, differential conductance \cite{Park08}, inelastic
neutron scattering \cite{Stock08}, and scanning tunneling microscopy
\cite{Allan13, Zhou13}, have discovered considerable evidence for a
$d_{x^2-y^2}$-wave superconducting gap. Angle-resolved in-plane
$H_{c2}$ has also been used to probe the gap symmetry of
CeCoIn$_{5}$. However, there is a longstanding experimental
discrepancy in the angular dependence of in-plane $H_{c2}$: some
experiments found that the maxima of $H_{c2}$ are along the [110]
direction \cite{Murphy02}, whereas other experiments observed the
maxima of $H_{c2}$ along the [100] direction \cite{Settai01,
Bianchi03, Weickert06}. This discrepancy is regarded as an open
puzzle in this field \cite{Weickert06, Das13}, and prevents us from
reaching a final consensus on the precise gap symmetry of
CeCoIn$_{5}$.

An external magnetic field couples to the charge and spin degrees of
freedom of electrons via the orbital and Zeeman mechanisms,
respectively. The former coupling destroys the long-range phase
coherence and leads to the mixed state in type-II superconductors.
The latter one, called Pauli paramagnetic effect, is believed to
play an important role in heavy fermion compounds such as
CeCu$_{2}$Si$_2$ and CeCoIn$_5$ \cite{Stockert12, Steglich79,
Weickert06, Bianchi02, Bianchi08}. The behavior of $H_{c2}$ is
determined by the interplay of these two effects.

It is well established that the in-plane $H_{c2}$ exhibits a
fourfold oscillation in $d$-wave superconductors \cite{Won94,
Takanaka95, Koike96, Naito01, Won04, Weickert06, Vieyra11}. In
earlier calculations including only the orbital effect \cite{Won94,
Takanaka95}, $H_{c2}$ was always found to display its maxima along
the antinodal directions where the $d$-wave gap is maximal. Later
studies included the Pauli paramagnetic effect \cite{Weickert06,
Vorontsov10}, but still concluded that the maxima of $H_{c2}$ are
along the antinodal directions. There appears to be \emph{a priori}
hypothesis in the literature that a larger superconducting gap
necessarily results in a larger $H_{c2}$, which means that $H_{c2}$
and $d$-wave gap should exhibit their maxima (minima) at exactly the
same azimuthal angles $\theta$. If this hypothesized correspondence
is valid, it would be easy to identify the gap symmetry: the gap is
$d_{x^2-y^2}$-wave when the measured $H_{c2}$ exhibits its maxima
along the [100] direction; the gap is $d_{xy}$-wave when the
measured $H_{c2}$ exhibits its maxima along the [110] direction.

We emphasize that the above hypothesized connection between in-plane
$H_{c2}$ and $d$-wave gap, though intuitively reasonable, is
actually not always correct. If there is only orbital effect,
$H_{c2}$ and $d$-wave gap do display the same angular dependence.
However, this connection can be destroyed by the Pauli paramagnetic
effect.

In this paper, motivated by the recent progress and the existing
controversy, we will analyze the influence of the interplay of
orbital and Pauli paramagnetic effect on the behavior of in-plane
$H_{c2}$ in $d$-wave superconductors. The aim of this paper is to
provide a better understanding of the properties of the angle-resolved
in-plane $H_{c2}$ in CeCu$_{2}$Si$_{2}$ and CeCoIn$_{5}$.
After carrying out systematical calculations, we show that the
maxima of angle-dependent $H_{c2}(\theta)$ are not necessarily along
the antinodal directions in the presence of Pauli paramagnetic
effect. Interestingly, the angular dependence of $H_{c2}$ is
determined by a number of parameters, including temperature $T$,
critical temperature $T_{c}$, gyromagnetic ratio $g$, fermion
velocity $v_0$, and two parameters that characterize the shape of
the corresponding Fermi surface. Any of these six parameters can
drive a $\pi/4$ shift in the fourfold oscillation pattern of
$H_{c2}$. Since approximations are inevitable in theoretical
calculations, it is technically quite difficult to identify whether
the precise gap symmetry is $d_{x^2-y^2}$ wave or $d_{xy}$ wave by
fitting experimental results with model calculations at a fixed
temperature. Among the six tuning parameters, the temperature $T$
plays a particular role. If one varies $T$ but fixes all the rest
parameters, $H_{c2}$ exhibits its maxima along the nodal directions
at $T < T^{*}$ and antinodal directions at $T > T^{*}$ due to a
sufficiently strong Pauli paramagnetic effect, where $T^{*}$ is
certain critical temperature. The concrete magnitude of $T^{*}$ may
change as other parameters vary, but the existence of a $\pi/4$
shift in the four-fold oscillation of $H_{c2}$ at $T^{*}$ appears to
be general feature. This feature provides a better method to
determine the precise gap symmetry by measuring the in-plane
$H_{c2}$ at a large number of temperatures and see whether there is
a $\pi/4$ shift in its angular dependence.

On the basis of our theoretical results, we find that some previous
conclusions about the precise gap symmetry of CeCu$_{2}$Si$_{2}$ are
actually unreliable, and need to be further studied. Moreover, our
finding provides a possible solution for an experimental discrepancy
in the measured angular dependence of in-plane $H_{c2}$ in
CeCoIn$_{5}$.

To examine the validity of our conclusion, we will also study the
impacts of Landau level mixing and Fulde-Ferrell-Larkin-Ovchinnikov
(FFLO) state \cite{Fulde64, Larkin64}, which may be important in
some heavy fermion compounds. We find that Landau level mixing does
not change the general feature that the maxima of $H_{c2}$ shifts by
$\pi/4$ at critical temperature $T^{*}$. In the presence of FFLO
state, however, the maximum of $H_{c2}$ may be along the nodal or
antinodal directions, depending sensitively on the temperature of
the system. This makes it more difficult to identify the precise gap
symmetry by measuring the angular dependence of $H_{c2}$.

The rest of paper is organized as follows. In
Sec.~\ref{Sec:Derivation}, we derive the equation of in-plane
$H_{c2}$ for $d_{x^2-y^2}$ superconductor with a rippled cylindrical
Fermi surface. In Sec.~\ref{Sec:ResultsExpCom}, we show the
influence of different parameters on $H_{c2}$ by numerical
calculations. In Sec.~\ref{Sec:LLAndFFLO}, the influences of Landau
level mixing and FFLO state on $H_{c2}$ are given. In
Sec.~\ref{Sec:CompExpe}, we then compare our results with
experimental studies about angle-resolved $H_{c2}$. In
Sec.~\ref{Sec:SumDis}, we summarize our main results.

\section{Derivation of the equation of $H_{c2}$\label{Sec:Derivation}}

\begin{figure}[htbp]
\center
\includegraphics[width=2.6in]{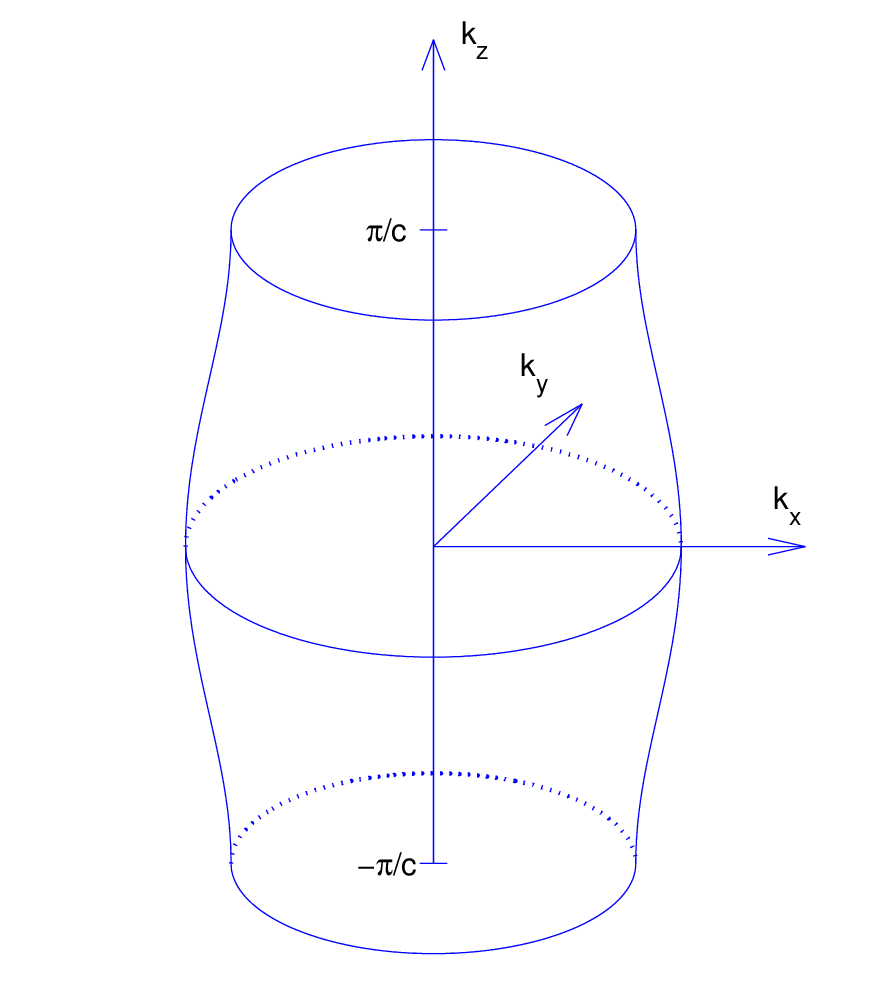}
\caption{Schematic diagram for a rippled cylindrical Fermi
surface.}\label{Fig:FS}
\end{figure}

Many heavy fermion compounds have a layered structure, but the
inter-layer coupling cannot be entirely ignored \cite{Matsuda06,
Sarrao07, Chubukov}. To embody this feature, it is convenient to
assume a rippled cylindrical Fermi surface, which is schematically
shown in Fig.~\ref{Fig:FS}. Now the fermion momentum $\mathbf{k}$
should have three components: $k_{x,y}$ denote the $x,y$-components
in the basic superconducting plane, and $k_z$ denotes the
$z$-component along $c$-axis. We use $t_{c}$ to represent the
inter-layer hoping parameter and $c$ the unit size along
$z$-direction. The dispersion relation of fermions is given by
\cite{Thalmeier05, Vorontsov07a},
\begin{eqnarray}
\varepsilon(\mathbf{\mathbf{k}}) =
\frac{1}{2m}(k_{x}^{2}+k_{y}^{2}) - 2t_{c}\cos(\chi)
\end{eqnarray}
with $\chi = k_{z}c$. Superconductivity is entirely suppressed once
the in-plane field $H$ reaches $H_{c2}$, which can be obtained by
solving a linearized gap equation. Near the second-order transition,
the gap function has the form
\begin{eqnarray}
\Delta\left(\mathbf{R},\mathbf{k}\right) =
\Delta_{\alpha}(\mathbf{R})\gamma_{\alpha}(\hat{\mathbf{k}}),
\end{eqnarray}
where $\gamma_{\alpha}(\hat{\mathbf{k}})$ reflects the symmetry of
the gap function and $\alpha$ may correspond to $s$,
$d_{x^2-y^2}$, $d_{xy}$, and so on. Employing the general methods
presented in Refs.~\cite{Helfand66, Werthamer66, Scharnberg80,
Lukyanchuk87, Shimahara96, Shimahara97, Suginishi06, Shimahara09},
we find the following equation
\begin{eqnarray}
-\ln(\frac{T}{T_{c}})\Delta_{\alpha}(\mathbf{R}) &=&
\int_{0}^{+\infty}d\eta\frac{\pi T}{\sinh(\pi T\eta)}
\int_{-\pi}^{\pi}\frac{d\chi}{2\pi}
\int_{0}^{2\pi}\frac{d\varphi}{2\pi}
\nonumber \\
&& \times \gamma_{\alpha}^2(\hat{\mathbf{k}})\left\{1-
\cos\left[\eta\left(h' + \frac{1}{2}
\mathbf{v}_{F}(\hat{\mathbf{k}})
\right.\right.\right. \nonumber \\
&&\left.\left.\left.\cdot\mathbf{\Pi}(\mathbf{R})\right)\right]
\right\}\Delta_{\alpha}(\mathbf{R}). \label{eqn:GapL}
\end{eqnarray}
In the simplest case, we now neglect the influence of Landau level
mixing and FFLO state. Their influence will be considered separately
in Sec.\ref{Sec:LLAndFFLO}. Assuming that
$\Delta_{\alpha}(\mathbf{R}) = \Delta_{0}(\mathbf{R})$, we have
\begin{eqnarray}
\Delta_{0}(\mathbf{R}) = \left(\frac{2eH}{\pi}\right)^{\frac{1}{4}}
e^{-eH\left(x\sin\theta-y\cos\theta\right)^{2}},
\end{eqnarray}
where $\Delta_{0}(\mathbf{R})$ is the lowest Landau level, and
$\theta$ is the angle between the direction of in-plane magnetic
field and the $x$-axis, corresponding to the [100] direction, within
the basal plane. The generalized derivative operator is defined as
\begin{eqnarray}
\mathbf{\Pi}(\mathbf{R}) =
-i\mathbf{\nabla}_{\mathbf{R}} + 2e\mathbf{A}(\mathbf{R}),
\end{eqnarray}
where the vector potential is chosen to be
\begin{eqnarray}
\mathbf{A}(\mathbf{R}) = H\left(-x\sin\theta +
y\cos\theta\right)\mathbf{e}_{z}.
\end{eqnarray}
Now the field $\mathbf{H}$ takes the form
\begin{eqnarray}
\mathbf{H} &=& \mathbf{\nabla}\times\mathbf{A} =
H\cos\theta\mathbf{e}_{x}+H\sin\theta\mathbf{e}_{y}.
\end{eqnarray}

For a rippled cylindrical Fermi surface, the vector of Fermi
velocity is given by \cite{Thalmeier05}
\begin{eqnarray}
\mathbf{v}_{F}(\hat{\mathbf{k}}) = v_{a}\cos\varphi\mathbf{e}_{x}+
v_{a}\sin\varphi\mathbf{e}_{y}+ v_{c}\sin\chi\mathbf{e}_{z}.
\end{eqnarray}
Here, $v_{c} = 2t_{c}c$, and $v_{a} = v_{0} \sqrt{1 +
\lambda\cos(\chi)}$, where $v_{0} = \frac{k_{F0}}{m}$ with Fermi
momentum $k_{F0}$ being related to Fermi energy $\epsilon_{F}$ by
$k_{F0} = \sqrt{2m\epsilon_{F}}$. The shape of rippled cylindrical
Fermi surface is characterized by a velocity ratio $v_{c}/v_{0} =
\lambda\gamma$, where $\lambda = 2t_{c}/\epsilon_{F}$ and $\gamma =
ck_{F0}/2$. As will be shown later, both $\lambda$ and $\gamma$ can
strongly affect the behavior of $H_{c2}$. Moreover, we define $h' =
-\frac{g\mu_{B}H}{2}$, where $\mu_B$ is Bohr magneton and $g$
gyromagnetic ratio. The orbital effect is encoded in the factor
$\mathbf{v}_{F}(\mathbf{k}) \cdot \Pi(\mathbf{R})$, whereas the
Pauli paramagnetic effect is represented by the factor $h'$. The
concrete angular dependence of $H_{c2}$ is determined by the
interplay of these two effects.

To facilitate analytical computation, we can choose the direction of
field $\mathbf{H}$ as a new $z'$-axis and define
\begin{eqnarray}
\left\{
\begin{array}{l}
\mathbf{e}_{x}'=\mathbf{e}_{x}\sin\theta-\mathbf{e}_{y}\cos\theta
\\
\mathbf{e}_{y}'=-\mathbf{e}_{z}
\\
\mathbf{e}_{z}'=\mathbf{e}_{x}\cos\theta+\mathbf{e}_{y}\sin\theta
\end{array}\right..
\end{eqnarray}
In the coordinate frame spanned by $(\mathbf{e}_{x}',
\mathbf{e}_{y}', \mathbf{e}_{z}')$, we write the velocity vector as
\begin{eqnarray}
\mathbf{v}_{F}(\hat{\mathbf{k}}) = v_{a}\sin(\theta-\varphi)
\mathbf{e}_{x}' - v_{c}\sin\chi \mathbf{e}_{y}' + v_{a}\cos(\theta -
\varphi)\mathbf{e}_{z}', \nonumber
\label{eqn:FermionV2}
\end{eqnarray}
and the generalized derivative operator as
\begin{eqnarray}
\mathbf{\Pi}(\mathbf{R}) &=& \sqrt{eH}
\left[\left(a_{+}+a_{-}\right)\mathbf{e}_{x}' -
i\left(a_{+}-a_{-}\right)\mathbf{e}_{y}' \right.\nonumber
\\
&&\left.+\sqrt{2}a_{0}\mathbf{e}_{z}'\right], \label{eqn:PiDef}
\end{eqnarray}
where
\begin{eqnarray}
a_{\pm} &=& \frac{1}{2\sqrt{eH}}\left[-i\sin\theta\partial_{x} +
i\cos\theta\partial_{y} \mp \partial_{z}\right. \nonumber \\
&&\left.\pm 2ieH(x\sin\theta-y\cos\theta)\right],
\label{Eq:RaiseReduceOperator}\\
a_{0} &=& \frac{1}{\sqrt{2eH}}\left[-i\partial_{x}\cos\theta -
i\partial_{y}\sin\theta \right],
\end{eqnarray}
which satisfy
\begin{eqnarray}
[a_{-},a_{+}] = 1, [a_{\pm},a_{0}] = 0.
\end{eqnarray}

\begin{figure}[htbp]
\includegraphics[width=3.4in]{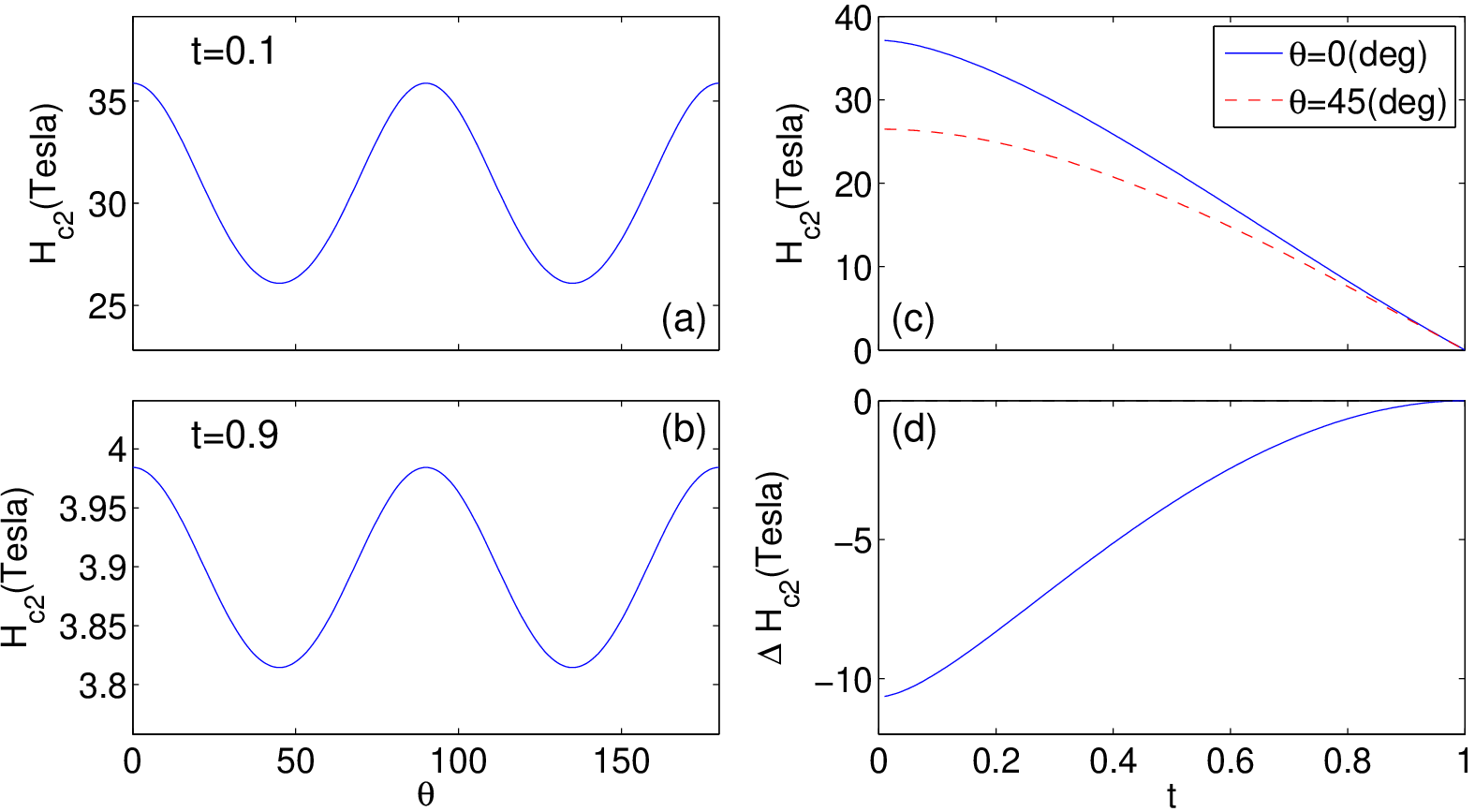}
\caption{(a) and (b): Fourfold oscillation of $\theta$-dependent
$H_{c2}$ at $t = 0.1$ and $t = 0.9$; (c) and (d): $t$-dependence of
$H_{c2}$ and $\Delta H_{c2}$ with $T_{c}=1K$, $v_{0} = 3000m/s$,
$\lambda = 0.5$, and $\gamma=1$. Only orbital effect is considered.}
\label{Fig:Hc2tNP}
\end{figure}

In Eq.~(\ref{eqn:GapL}), the influence of gap symmetry is reflected
in $\gamma_{\alpha}(\hat{\mathbf{k}})$. For $s$-wave gap,
$\gamma_{s}(\hat{\mathbf{k}}) = 1$; for $d_{x^2 - y^2}$-wave gap,
$\gamma_{d_{x^2-y^2}}(\hat{\mathbf{k}}) = \sqrt{2}\cos(2\varphi)$;
for $d_{xy}$-wave gap, $\gamma_{d_{xy}}(\hat{\mathbf{k}}) =
\sqrt{2}\sin(2\varphi)$. Now we take $d_{x^2-y^2}$-wave gap as an
example, so $\gamma_{d_{x^2-y^2}}(\hat{\mathbf{k}}) =
\sqrt{2}\cos(2\varphi)$. The results for $d_{xy}$-wave gap can be
obtained analogously, and the main conclusion does not change.
Averaging over $\Delta(\mathbf{R})$ on both sides of
Eq.~(\ref{eqn:GapL}) and inserting
$\gamma_{d_{x^2-y^2}}(\hat{\mathbf{k}})$, we obtain
\begin{eqnarray}
-\ln t &=& \int_{0}^{+\infty}\frac{du}{\sinh\left(u\right)}
\left\{1 - \int_{-\pi}^{\pi}\frac{d\chi}{2\pi}
\int_{0}^{2\pi}\frac{d\varphi}{2\pi}\right. \nonumber \\
&&\times \cos(h u) \left[1+\cos(4\theta)\cos(4\varphi)\right]
\nonumber \\
&&\times \left. e^{-\rho u^2
\left(\lambda^2\gamma^2\sin^2\chi +
\left(1+\lambda\cos\chi\right)\sin^{2}\varphi \right)}\right\},
\label{Eq:Hc2Expression}
\end{eqnarray}
where $t=\frac{T}{T_{c}}$, $h=\frac{g\mu_{B}H_{c2}}{2\pi k_{B}T}$
and $\rho=\frac{v_{0}^{2}eH_{c2}}{8\pi^2 k_{B}^{2}T^2}$. For the
detailed derivation of Eq.~(\ref{Eq:Hc2Expression}), please see the
Appendix.

Although the linearized gap equation (\ref{Eq:Hc2Expression}) is
formally general and valid in many superconductors, its solution is
determined by a number of physical effects and associated
parameters. From Eq.~(\ref{Eq:Hc2Expression}), we see
$H_{c2}(\theta)$ depends on six physical parameters: critical
temperature $T_{c}$, temperature ratio $t=T/T_{c}$, velocity
$v_{0}$, gyromagnetic ratio $g$, $\lambda=2t_{c}/\epsilon_{F}$, and
$\gamma = k_{F0}c/2$. Among these parameters, $\lambda$ and $\gamma$
are related to the shape of the rippled cylindrical Fermi surface.
We notice that the influence of $\lambda$ and $\gamma$ were not
carefully investigated in previous works on $H_{c2}$.
x

\begin{figure}[htbp]
\includegraphics[width=2.5in]{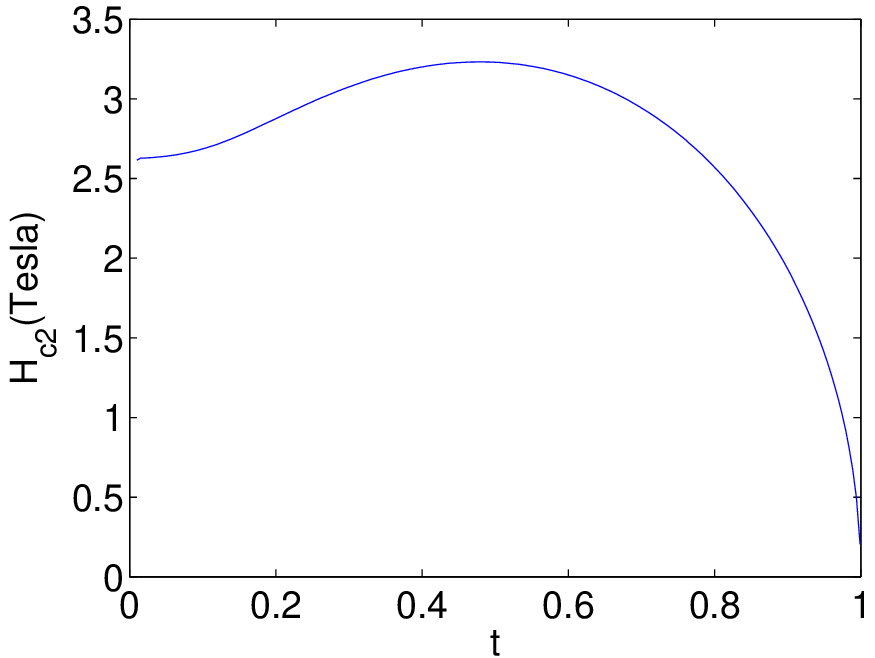}
\caption{$t$-dependence of $H_{c2}$ caused solely by Pauli
paramagnetic effect. $T_{c}=1K$, $\lambda=0.5$, $\gamma = 1$, and $g
= 1$.} \label{Fig:PauliOnly}
\end{figure}

In previous studies on this problem \cite{Won04, Weickert06,
Vieyra11}, a rippled cylindrical Fermi surface is often assumed, but
there is not any tuning parameter in the equation of $H_{c2}$ that
can characterize how rippled is the Fermi surface. In our equation
of $H_{c2}$, given by Eq.~(\ref{Eq:Hc2Expression}), we have
introduced two tuning parameters $\lambda$ and $\gamma$ to
characterize the concrete shape of the rippled Fermi surface. In the
next section, we will show that whether the maxima of $H_{c2}$ is
along the nodal or antinodal direction depends on the specific
values of these two parameters. Apparently, the shape of the Fermi
surface can significantly influence the angular dependence of
$H_{c2}$, which is not properly considered in previous works. In
addition, in previous studies of $H_{c2}$ \cite{Won04, Weickert06,
Vieyra11}, the precise $d$-wave gap symmetry is identified by
comparing theoretical calculations to experimental results of
$H_{c2}$ at a fixed temperature. In the next section, we will prove
that varying the temperature leads to a $\pi/4$ shift of
angle-resolved $H_{c2}$. This striking temperature dependence of
angle-resolved $H_{c2}$ has not been realized previously. According
to this property, measuring $H_{c2}$ and then fitting experiments
with model calculations at a given temperature may yield incorrect
conclusion about the precise gap symmetry.

\section{Angular dependence of $H_{c2}$ and its connection with gap symmetry
\label{Sec:ResultsExpCom}}

In this section, we present the numerical results for in-plane
$H_{c2}$ by solving Eq.~(\ref{Eq:Hc2Expression}) numerically and
discuss the influence of various parameters on the angular
dependence of $H_{c2}$.

The detailed behavior of $H_{c2}$ can be clearly seen from its
angular dependence. In addition, it is equally important to analyze
the difference of $H_{c2}$ between its values obtained at $\theta =
45^{\degree}$ and $\theta = 0^{\degree}$, i.e., $\Delta H_{c2} =
H_{c2}(\theta = 45^{\degree}) - H_{c2}(\theta = 0^{\degree})$ since
the maxima and minima of $H_{c2}$ always appear at these two angles.
$H_{c2}$ exhibits its maxima at $\theta = 45^{\degree}$ if $\Delta
H_{c2} > 0$ and at $\theta = 0^{\degree}$ if $\Delta H_{c2} < 0$.

First, we consider only the orbital effect by setting $g = 0$. In
this case, the factor $\cos(hu)$ appearing in
Eq.~(\ref{Eq:Hc2Expression}) is equal to unity, $\cos(hu) = 1$. We
assume that $T_c = 1K$, $v_{0} = 3000m/s$, $\lambda=0.5$, and
$\gamma=1$, which are suitable parameters for heavy fermion
compounds.

\begin{figure}[htbp]
\includegraphics[width=3.4in]{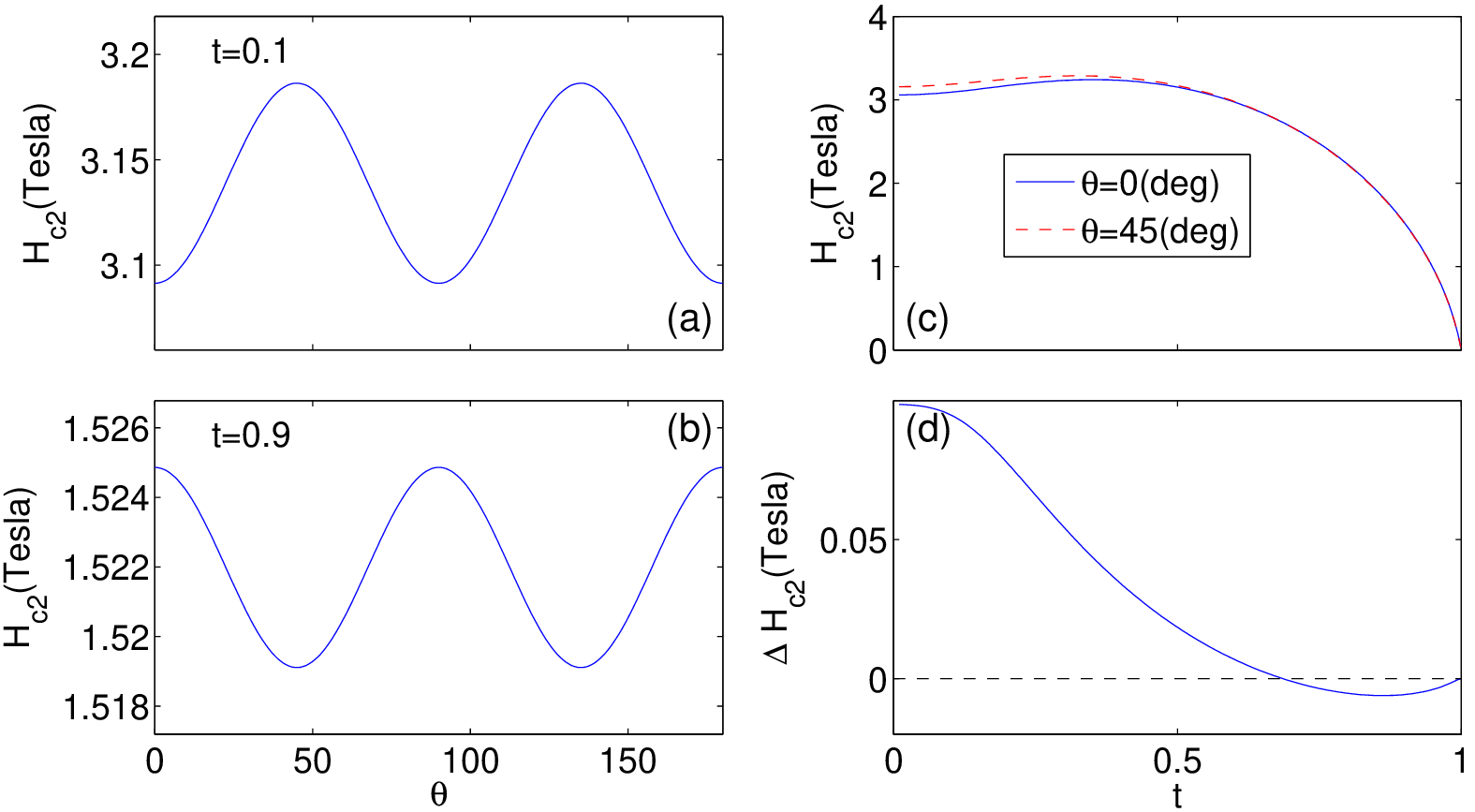}
\caption{(a) and (b): Angular dependence of $H_{c2}$ at $t = 0.1$
and $t = 0.9$. $T_{c} = 1K$, $v_{0} = 3000m/s$, $\lambda=0.5$,
$\gamma = 1$, and $g=1$. (c) and (d): $t$-dependence of $H_{c2}$ and
$\Delta H_{c2}$. Both the orbital and Pauli effects are considered.}
\label{Fig:Hc2t}
\end{figure}

After carrying out numerical calculations, we plot the angular
dependence of $H_{c2}(\theta)$ in Fig.~\ref{Fig:Hc2tNP} at two
representative temperatures $t = 0.1$ and $t = 0.9$.
Figure~\ref{Fig:Hc2tNP} clearly shows that $H_{c2}(\theta)$ exhibits
a fourfold oscillation pattern. The maxima of $H_{c2}$ is always
along the antinodal directions for any values of relevant
parameters, which means that the angular dependence of orbital
effect-induced $H_{c2}$ is exactly the same as that of the gap. This
is consistent with previous results of Refs.~\cite{Won94,
Takanaka95}. Moreover, the positions of peaks are $t$-independent.
$H_{c2}$ is a monotonic decreasing function of $t$, since the gap is
suppressed as $t$ grows. Moreover, $\Delta H_{c2}$ is negative for
all values of $t$.

We next consider the influence of pure Pauli paramagnetic effect on
$H_{c2}$ by setting $v_{0}=0$, which leads to
\begin{eqnarray}
-\ln(t) = \int_{0}^{+\infty}du\frac{1 - \cos(hu)}{\sinh(u)}.
\end{eqnarray}
This equation is completely independent of $\theta$. The
$t$-dependence of $H_{c2}$ is shown in Fig.~\ref{Fig:PauliOnly}.
$H_{c2}$ is not a monotonic function: it first rises with growing
$t$, but decreases when $t$ is sufficiently large.

Finally we come to the interplay of orbital and Pauli paramagnetic
effects, which are both important in some heavy fermion compounds,
including CeCu$_{2}$Si$_2$ and CeCoIn$_5$. As aforementioned, the
angular dependence of $H_{c2}$ is determined by a number of tuning
parameters. To make the results as transparent as possible, we vary
one single parameter at each time and fix all the rest parameters at
certain values.

As shown in Fig.~\ref{Fig:Hc2t}, under the chosen parameters, the
maxima of $H_{c2}$ locates along the antinodal directions at a
relatively higher temperature $t = 0.9$. This behavior is very
similar to that in the case of pure orbital effect. However, at a
lower temperature $t = 0.1$, the maxima of $H_{c2}$ is along the
nodal directions where the $d_{x^2 - y^2}$-wave gap vanishes. Two
conclusions can be drawn: (\romannumeral1) $H_{c2}$ does not always
exhibit its maxima at the angles where the superconducting gap is
maximal; (\romannumeral2) the fourfold oscillation curves of
$H_{c2}$ are shifted by $\pi/4$ as temperature grows across certain
critical value $T^{*}$.

We see from Fig.~\ref{Fig:Hc2t}(c) that $H_{c2}$ first increases
with growing $t$, but decreases rapidly once $t$ exceeds a critical
value. Such a non-monotonic $t$-dependence of $H_{c2}$ is clearly
caused by the Pauli paramagnetic effect. Moreover, the difference
$\Delta H_{c2}$ shown in Fig.~\ref{Fig:Hc2t}(d) is positive for
small $t$ but becomes negative for larger values of $t$.

\begin{figure}[htbp]
\includegraphics[width=3.4in]{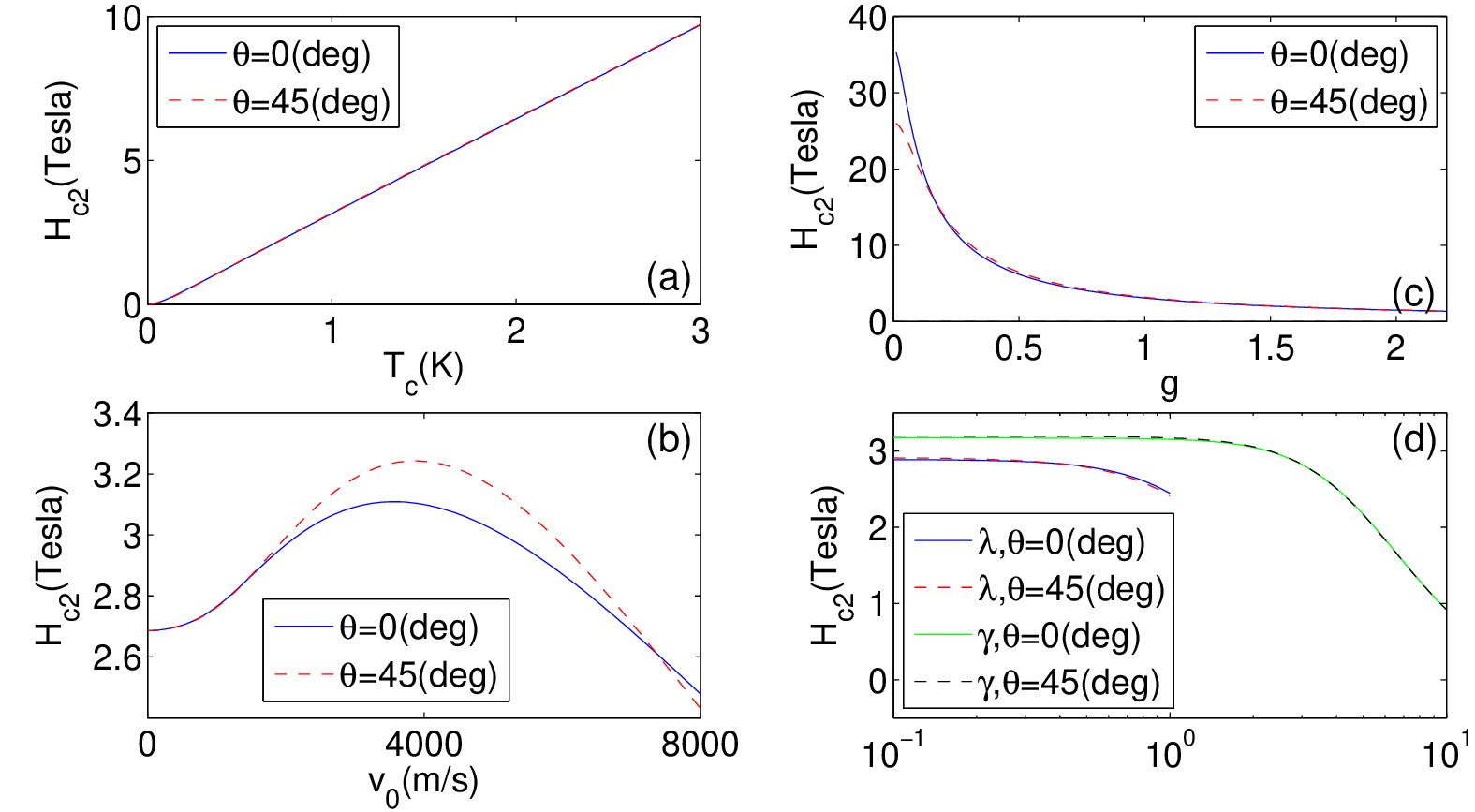}
\caption{(a) $T_c$-dependence of $H_{c2}$. $t = 0.5$, $v_{0} =
3000m/s$, $\lambda=0.5$, $\gamma=1$, and $g=1$. (b)
$v_{0}$-dependence of $H_{c2}$. $t=0.1$, $T_{c}=1K$, $\lambda=0.5$,
$\gamma=1$, and $g=1$. (c) $g$-dependence of $\Delta H_{c2}$. $t =
0.1$, $v_{0} = 3000m/s$, $T_{c} = 1K$, $\lambda = 0.5$, and $\gamma
= 1$. (d) Blue solid line and red dashed line: $\lambda$-dependence
of $H_{c2}$ for $\theta=0^{\degree}$ and $\theta=45^{\degree}$
respectively. $t = 0.5$, $v_{0} = 5000m/s$, $T_{c} = 1K$,
$\gamma=1$, and $g = 1$; Green solid line and black dashed line:
$\gamma$-dependence of $\Delta H_{c2}$ for $\theta=0^{\degree}$ and
$\theta=45^{\degree}$ respectively. $t = 0.5$, $v_{0} = 3000m/s$,
$T_{c} = 1K$, $\lambda =0.5$, and $g = 1$. Both the orbital and
Pauli effects are considered.} \label{Fig:Hc2Both}
\end{figure}

\begin{figure}[htbp]
\includegraphics[width=3.4in]{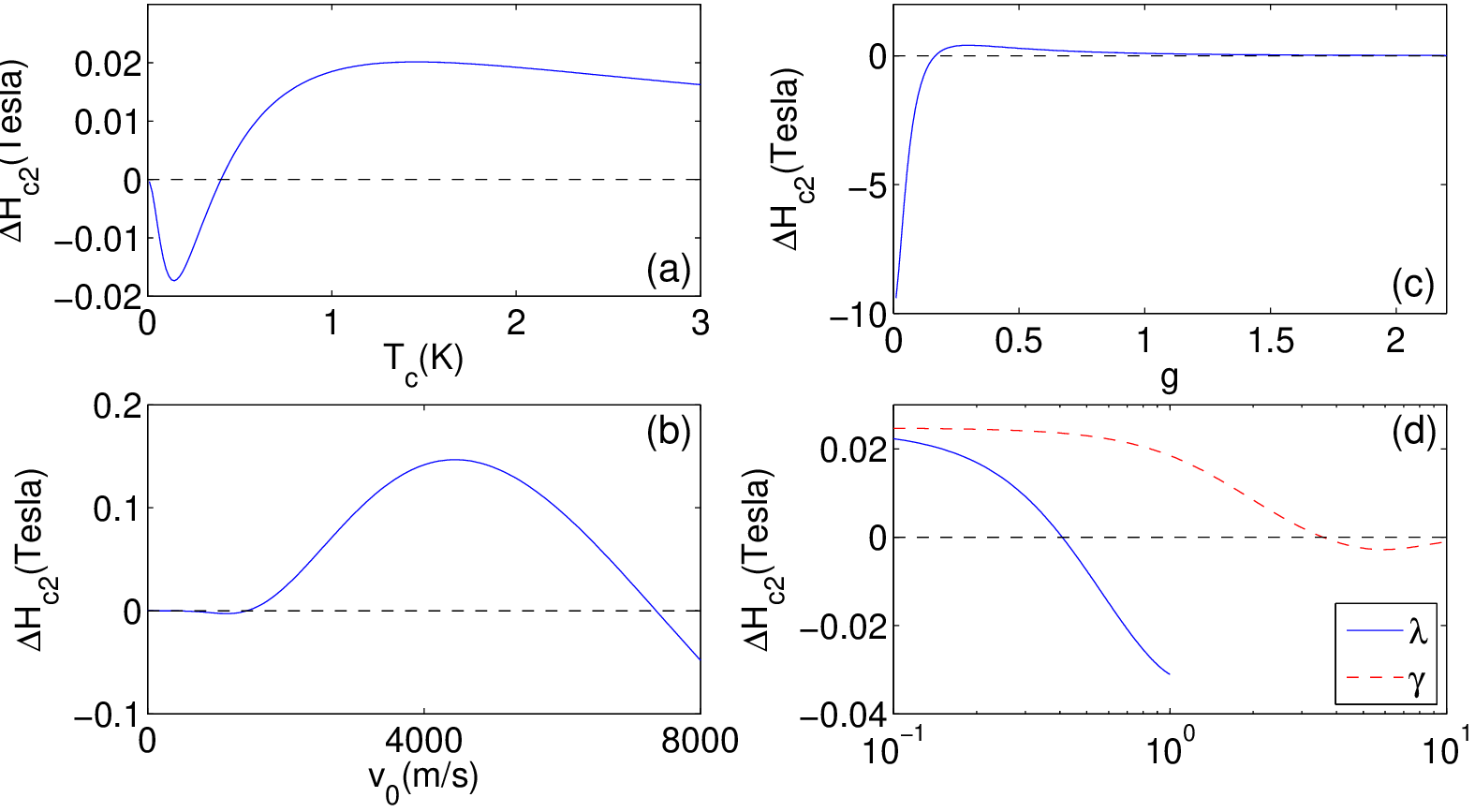}
\caption{(a) $T_c$-dependence of $\Delta H_{c2}$.  (b)
$v_{0}$-dependence of $\Delta H_{c2}$. (c) $g$-dependence of $\Delta
H_{c2}$. (d) Blue solid line: $\lambda$-dependence of $\Delta
H_{c2}$.  Red dashed line: $\gamma$-dependence of $\Delta H_{c2}$.
Parameters are the same as those of Fig.~\ref{Fig:Hc2Both}.}
\label{Fig:DeltaHc2}
\end{figure}

$T_c$: It is well known that $T_c$ of heavy fermion compounds is not
high, especially when compared with cuprates and iron pnictides. To
make a general analysis, we assume $T_{c}$ varies between 0K and 3K.
All the other parameters are fixed. From Fig.~\ref{Fig:Hc2Both}(a),
we find that $H_{c2}$ increases monotonously as $T_{c}$ grows.
As displayed in Fig.~\ref{Fig:DeltaHc2}(a), if $T_{c}$ is smaller
than some critical value, $\Delta H_{c2}$ is negative, which means
the maxima of $H_{c2}$ are along the antinodal directions. For
larger $T_{c}$, $\Delta H_{c2}$ becomes positive and the maxima of
$H_{c2}$ are shifted to nodal directions. Clearly, $T_c$ has
important impacts on the angular dependence of $H_{c2}$.

$v_0$: We then consider the influence of fermion velocity $v_0$,
which characterizes the strength of the orbital effect. According to
Fig.~\ref{Fig:Hc2Both}(b), $H_{c2}$ is not a monotonic function of
$v_{0}$, it increases with $v_{0}$ for small values of $v_{0}$ but
decreases with $v_{0}$ when $v_{0}$ is large enough. Therefore, in
the presence of of Pauli paramagnetic effect, the increasing of
orbital effect does not necessarily suppress $H_{c2}$. At $v_{0}
=0$, the orbital effect is removed, so the Pauli paramagnetic effect
entirely determines $H_{c2}$. $H_{c2}$ is then angle independent,
and $\Delta H_{c2} = 0$.  For finite $v_0$, $H_{c2}$ becomes angle
dependent and exhibits fourfold oscillation, as a consequence of the
interplay between orbital and Pauli paramagnetic effects. As shown
in Fig.~\ref{Fig:DeltaHc2}(b), $\Delta H_{c2}$ is negative for both
small and large values of $v_{0}$, but becomes positive for
intermediate values of $v_{0}$.

$g$: Taking $g = 0$ simply leads to the known results obtained in
the case of pure orbital effect. From Fig~\ref{Fig:Hc2Both}(c), we
see that $H_{c2}$ monotonously deceases with the increase of $g$,
which indicates the Pauli paramagnetic effect always tends to
suppress $H_{c2}$. As depicted in Fig.~\ref{Fig:DeltaHc2}(c),
$\Delta H_{c2}$ is negative when $g$ takes small values but positive
when $g$ becomes sufficiently large.

\begin{figure}[htbp]
\includegraphics[width=2.6in]{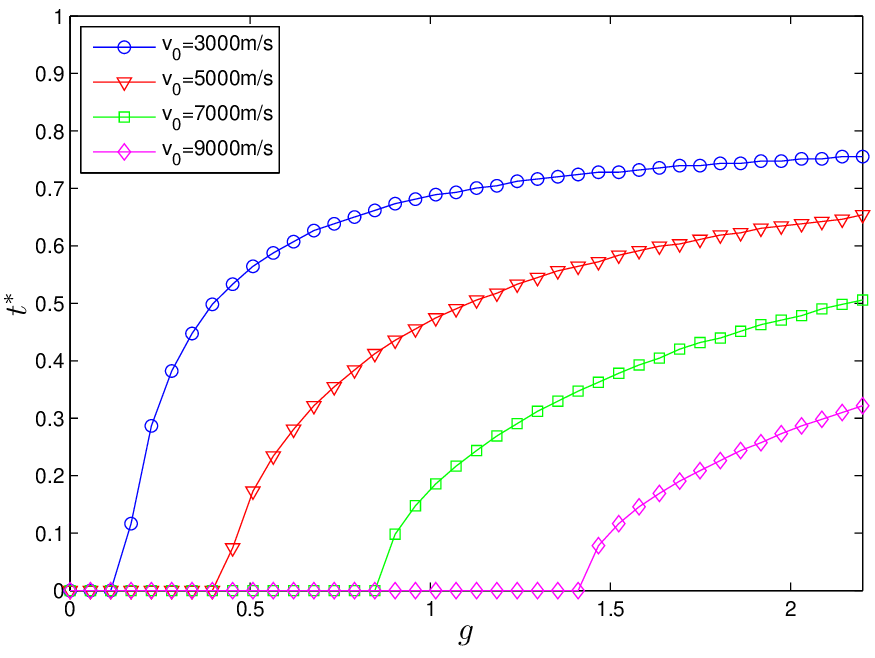}
\caption{Relation between $g$ and $t^{*}$. $T_{c} = 1K$, $\lambda =
0.5$, $\gamma = 1$.} \label{Fig:tStar}
\end{figure}

$\lambda$: For $t_{c}= \lambda = 0$, the rippled cylindrical Fermi
surface reduces to a cylindrical Fermi surface.
Figure~\ref{Fig:Hc2Both}(d) shows that $H_{c2}$ decreases
monotonously as $\lambda$ increases. It appearers that $H_{c2}$
takes larger values as a three-dimensional superconductor evolves
gradually to be quasi-two-dimensional. According to the blue solid
line in Fig.~\ref{Fig:DeltaHc2}(d), for given values of other
parameters, the maxima of $H_{c2}$ is along the nodal directions for
small values of $\lambda$ and antinodal directions for large values
of $\lambda$.

$\gamma$: As shown in Fig~\ref{Fig:Hc2Both}(d), $H_{c2}$ is a
monotonic function of $\gamma$. Varying $\gamma$ can also lead to
similar $\pi/4$ shift in $H_{c2}$. According to the red dashed line
in Fig.~\ref{Fig:DeltaHc2}(d), for given relevant parameters, the
maxima of $H_{c2}$ is along nodal directions for small values of
$\gamma$ and antinodal directions for large values of $\gamma$.

\begin{figure}[htbp]
\includegraphics[width=3.4in]{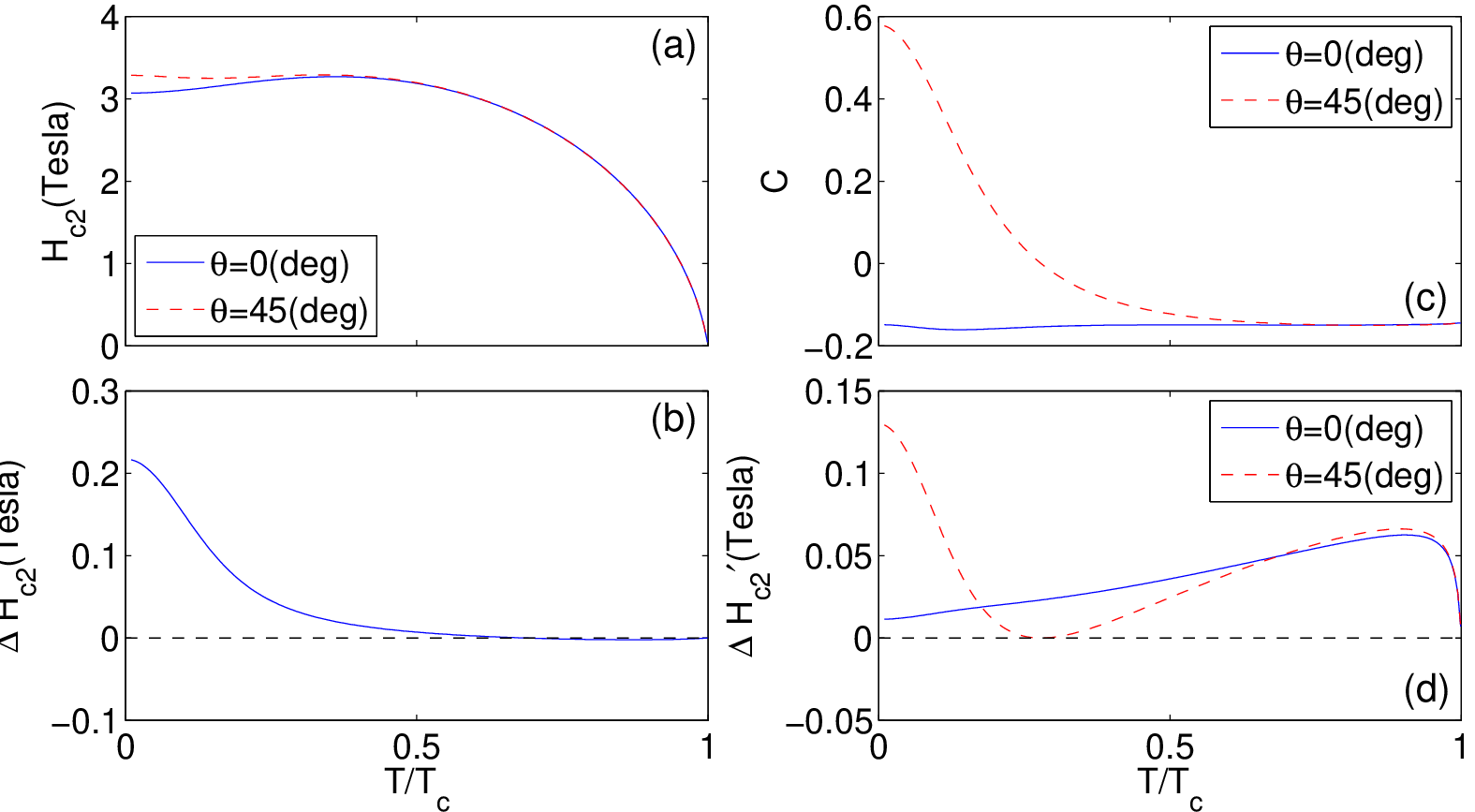}
\caption{(a), (b), (c) and (d): $t$-dependence of $H_{c2}$, $\Delta
H_{c2}$, $C$ and $\Delta H_{c2}'$. $T_{c} = 1K$, $v_{0} = 3000m/s$,
$\lambda=0.5$, $\gamma = 1$, and $g=1$. Landau level mixing is
included.} \label{Fig:Hc2tMixing}
\end{figure}

\begin{figure}[htbp]
\includegraphics[width=3.4in]{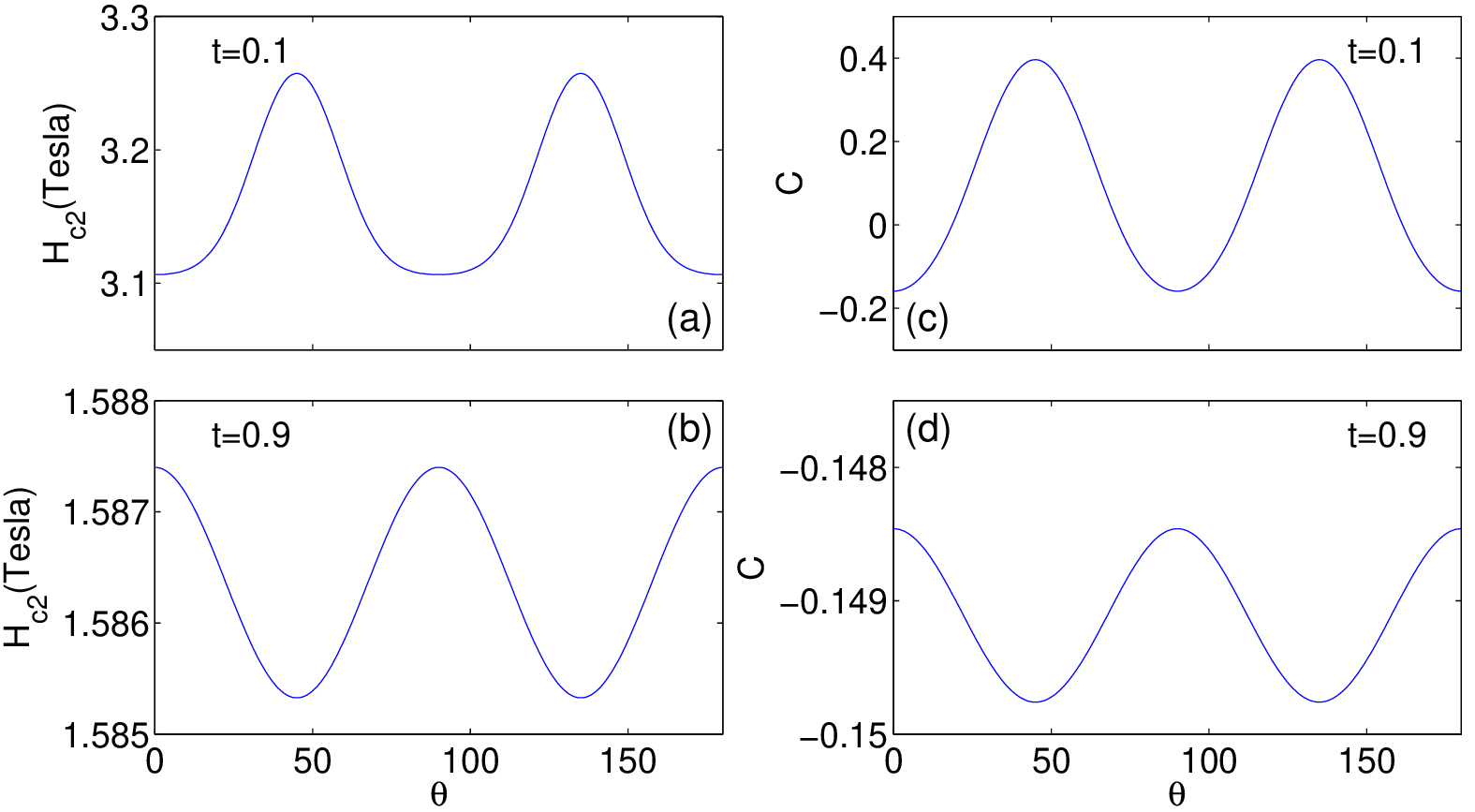}
\caption{Angular dependence of $H_{c2}$ and $C$ at $t = 0.1$ and $t
= 0.9$. $T_{c} = 1K$, $v_{0} = 3000m/s$, $\lambda=0.5$, $\gamma =
1$, and $g=1$. Landau level mixing is included}
\label{Fig:Hc2ThetaMixing}
\end{figure}

From above results, we know that the detailed angular dependence of
in-plane $H_{c2}$ is significantly influenced by a number of
physical parameters. The fourfold oscillation pattern of $H_{c2}$
can be shifted by $\pi/4$ if we tune anyone of these parameters. The
fact that $H_{c2}$ may exhibit its maxima along either nodal or
antinodal directions denies the naive expectation that $H_{c2}$
always displays the same angular dependence as the $d$-wave gap.
Therefore, one should be very careful when fitting theories with
experiments, because inaccurate and even wrong conclusions will be
drawn if some of the parameters are not properly chosen. Due to the
complicated dependence of $H_{c2}$ on various parameters and
inevitable approximations employed in the theoretical calculations,
it is infeasible to identify the precise gap symmetry solely by
measuring the fourfold oscillation of $H_{c2}$.

The temperature $t$ plays a particular role since it is usually the
only free parameter in one specific material. Our results show that
there is generally a $\pi/4$ difference in the positions of the
maxima of $H_{c2}$ and those of $d$-wave gap for $t <
t^{*}=T^{*}/T_{c}$, provided that $g$ is sufficiently large. We
emphasize that this conclusion does not depend on the specific
values of other five parameters. Indeed, those five parameters
change the fourfold oscillation of $H_{c2}$ by altering the critical
value $T^{*}$. However, $H_{c2}$ and $d$-wave gap always exhibit
their maxima at exactly the same angles once $t$ exceeds $t^{*}$,
which indicates that the Pauli paramagnetic effect is relatively
weak compared to the orbital effect at higher $T$.

In order to better understand this point, we plot the relation
between $t^{*}$ and $g$ for several values of velocity $v_{0}$ in
Fig.~\ref{Fig:tStar}. Since larger $g$ represents stronger Pauli
paramagnetic effect and larger $v_{0}$ describes stronger orbital
effect, this figure clearly shows how $t^{*}$ is determined by the
competition between the orbital and Pauli paramagnetic effects. The
monotonic increase of $t^{*}$ with growing $g$ confirms the
conclusion that strong Pauli paramagnetic effect causes the $\pi/4$
difference between the angular dependence of $H_{c2}$ and $d$-wave
gap.

\section{The influence of Landau level mixing and FFLO state\label{Sec:LLAndFFLO}}

In this section, we consider the influence of Landau level mixing
and the FFLO state. In contrast to the isotropic $s$-wave
superconductors, higher Landau level components of the order
parameter are generally mixed in anisotropic superconductors, which
was first emphasized by Luk'yanchuk and Mineev \cite{Lukyanchuk87,
Won94, Won96, Won04, Weickert06, Vieyra11}. In the case of $d$-wave
pairing, symmetry arguments ensure that only the $N=0$ and $N=2$
Landau levels are allowed \cite{Won96, Won04, Weickert06, Vieyra11}.
FFLO state is a novel superconducting state induced by strong
magnetic field where the corresponding Cooper paring has a finite
total momentum \cite{Fulde64, Larkin64, Shimahara09, Casalbuoni04,
Matsuda07}. In a FFLO state, the superconducting gap is modulated in
the real space. There have appeared considerable experimental clues
in the past decade suggesting that CeCoIn$_{5}$ is a possible
candidate for the FFLO state \cite{Thompson12, Bianchi03,
Matsuda07}.

Including the mixing between different Landau levels, the function
$\Delta_{\alpha}(\mathbf{R})$ can be written as \cite{Won94, Won04,
Weickert06, Vieyra11}
\begin{eqnarray}
\Delta_{\alpha}(\mathbf{R}) = \left[1+C(a_{+})^{2}\right]
\Delta_{0}(\mathbf{R}),
\end{eqnarray}
where $a_{+}$ is the raising operator which is showed in
Eq.~(\ref{Eq:RaiseReduceOperator}), and $C$ is the corresponding
admixing parameter of the Landau levels. The corresponding equations
for $H_{c2}$ is found to have the form
\begin{eqnarray}
-\ln(t) &=& \int_{0}^{+\infty}du\frac{1}{\sinh\left(u\right)}
\left\{1-\int_{-\pi}^{\pi}\frac{d\chi}{2\pi}
\int_{0}^{2\pi}\frac{d\varphi}{2\pi}\right.\nonumber
\\
&&\times\cos\left(hu\right)\left[1 +
\cos(4\varphi)\cos(4\theta)\right]\nonumber \\
&&\times\exp\left[-\rho u^2 S_{1}\right]\left[1+2C\rho
u^2S_{2}\right]\bigg\},\label{Eq:Hc2MixingA}
\end{eqnarray}
and
\begin{eqnarray}
-\ln(t)C &=& \int_{0}^{+\infty}du \frac{1}{\sinh\left(u\right)}
\left\{C - \int_{-\pi}^{\pi}\frac{d\chi}{2\pi}
\int_{0}^{2\pi}\frac{d\varphi}{2\pi}\right.\nonumber \\
&&\times\cos\left(hu\right)\left[1+\cos(4\varphi)
\cos(4\theta)\right]\nonumber \\
&&\times\exp\left[-\rho u^2 S_{1}\right] \left[\rho
u^2S_{2}\right.\nonumber \\
&&\left.+ C\left(1 - 4\rho u^2S_{1}
+2\rho^2u^4S_{1}^{2}\right)\right]\bigg\},\label{Eq:Hc2MixingB}
\end{eqnarray}
where
\begin{eqnarray}
S_{1} &=& \lambda^2\gamma^{2}\sin^{2}\chi +
\left(1+\lambda\cos\chi\right)\sin^{2}\varphi, \\
S_{2}&=&\lambda^2\gamma^2\sin^{2}\chi -
\left(1+\lambda\cos\chi\right)\sin^{2}\varphi.
\end{eqnarray}

\begin{figure}[htbp]
\includegraphics[width=2.5in]{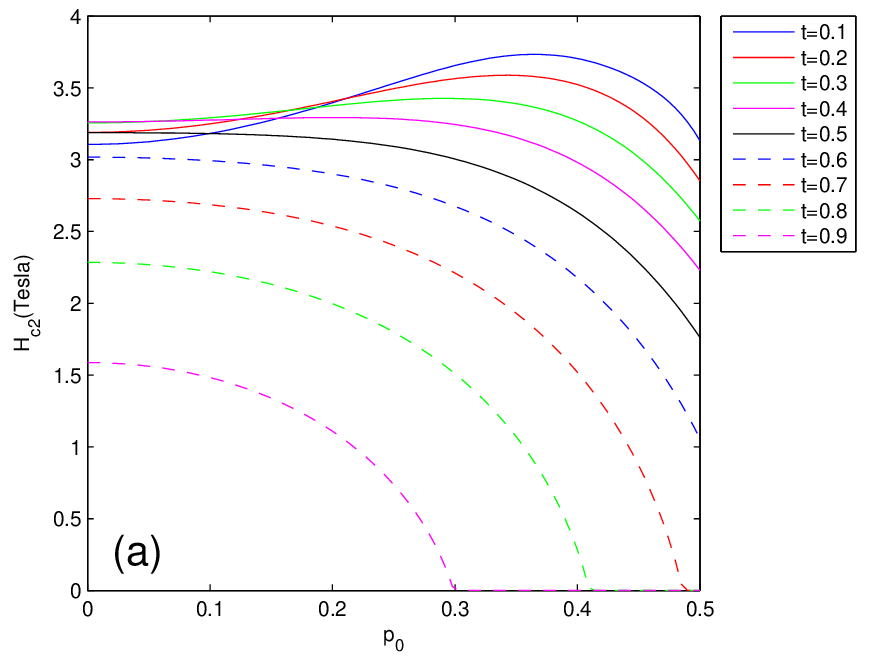}
\includegraphics[width=2.5in]{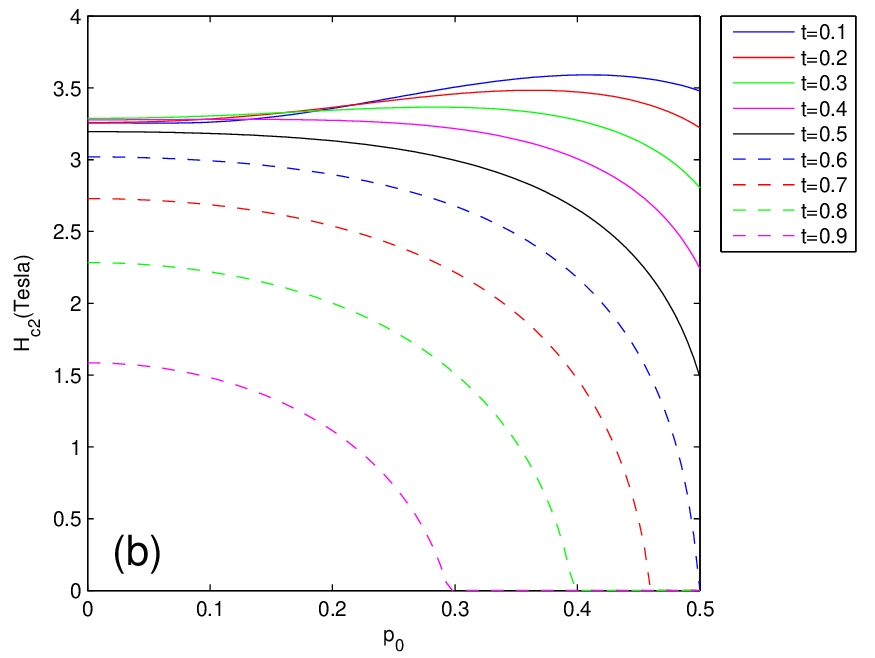}
\caption{The relation between $H_{c2}$ and the parameter $p_{0}$ at
different temperatures. The other parameters are chosen as $T_{c} =
1K$, $v_{0} = 3000m/s$, $\lambda=0.5$, $\gamma = 1$, and $g=1$. (a)
$\theta=0^{\degree}$; (b) $\theta=45^{\degree}$. Landau level mixing
and influence of FFLO state are included.} \label{Fig:Hc2FFLOCurve}
\end{figure}

We show the $T$-dependence of $H_{c2}$, $C$, $\Delta H_{c2}$, and
$\Delta' H_{c2}$ in Fig.~\ref{Fig:Hc2tMixing}, where $\Delta
H_{c2}'$ represents the difference of the values of $H_{c2}$ with
and without the Landau level mixing effects. The angular dependences
of $H_{c2}$ and $C$ are plotted in Fig.~\ref{Fig:Hc2ThetaMixing}. We
find that including Landau level mixing does not change the general
feature that the maximum of $H_{c2}$ is along nodal directions at
low $T$ but along antinodal directions at higher $T$.
Figure~\ref{Fig:Hc2tMixing} also shows that $\Delta H_{c2}'$ is
greater than zero, which simply means that the Landau level mixing
enhances $H_{c2}$.

We now consider the impacts of both Landau level mixing and FFLO
state. In this case, we should re-write
$\Delta_{\alpha}(\mathbf{R})$ as \cite{Weickert06}
\begin{eqnarray}
\Delta_{\alpha}(\mathbf{R}) = \cos(\mathbf{q}\cdot\mathbf{R})
\left[1+C(a_{+})^{2}\right]\Delta_{0}(\mathbf{R}).
\end{eqnarray}
The equations for $H_{c2}$ can be obtained by replacing the function
$\cos(hu)$ appearing in Eqs.~(\ref{Eq:Hc2MixingA}) and
(\ref{Eq:Hc2MixingB}) with
$\cos(hu)\cos\left[pu\cos(\varphi)\right]$, where $p =
\frac{v_{a}q}{2\pi T}$. A detailed derivation of the equations is
presented in Appendix \ref{Appendix:Hc2Derivation}.

\begin{figure}[htbp]
\includegraphics[width=3.4in]{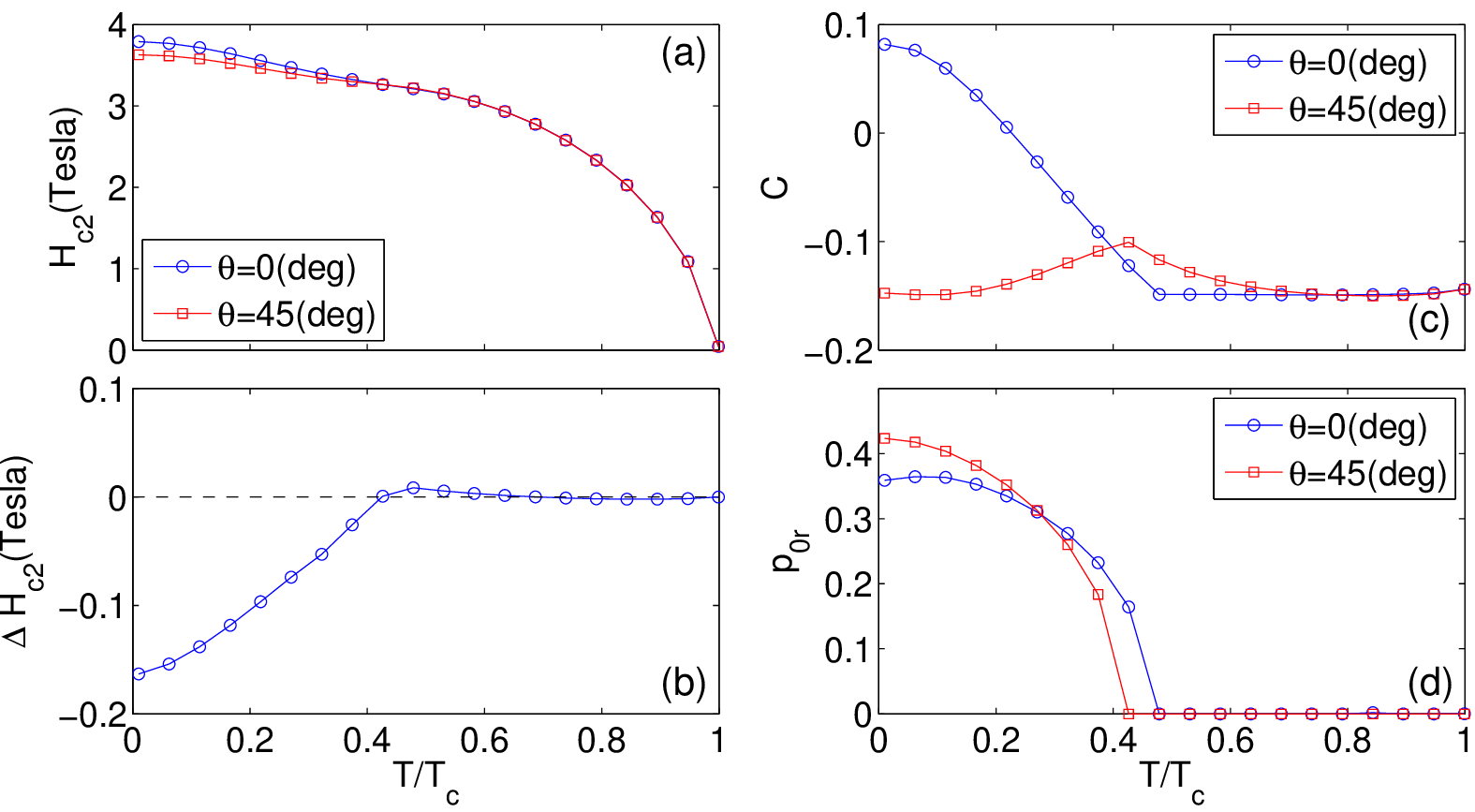}
\caption{Temperature dependence of $H_{c2}$, $\Delta H_{c2}$, $C$
and $p_{0r}$ are shown in (a), (b), (c) and (d) respectively. The
other parameters are chosen as $T_{c} = 1K$, $v_{0} = 3000m/s$,
$\lambda=0.5$, $\gamma = 1$, and $g=1$. Landau level mixing and
influence of FFLO state are included.} \label{Fig:Hc2FFLO}
\end{figure}

The relations between $H_{c2}$ and the FFLO parameter $p_{0}$ at
different temperatures are shown in Fig.~\ref{Fig:Hc2FFLOCurve}.
Here, $p_{0}$ is given by $p_{0} = \frac{v_{0}q}{2\pi T}$, and
$p_{0r}$ denotes the physical value of $p_{0}$ that corresponds to
the maximum value of $H_{c2}$. We find that $p_{0r}$ takes a finite
value at lower temperature, however, $p_{0r}$ equals to zero when
the temperature is larger than a critical value. After numerical
calculation, we find the temperature dependence of $H_{c2}$, $\Delta
H_{c2}$, $C$ and $p_{0r}$, as shown in Fig.~\ref{Fig:Hc2FFLO}. It is
interesting that, once FFLO state is considered, the angular
dependence of $H_{c2}$ can be significantly modified. We can see
that the maxima of $H_{c2}$ is along the antinodal directions at low
and high temperatures, but along the nodal directions at
intermediate temperatures. Apparently, the existence of FFLO state
makes it nearly impossible to probe the gap symmetry by measuring
$H_{c2}$ at some fixed temperature. Measuring the angular dependence
of $H_{c2}$ at a number of different temperatures is thus more
reasonable.

\section{Comparison with experiments\label{Sec:CompExpe}}

In this section, we compare our results with the experimental
studies. There is a longstanding controversy on the in-plane
$H_{c2}$ of CeCoIn$_{5}$. Settai \emph{et al.} \cite{Settai01}
reported that the maxima of $H_{c2}$ are along the [100] direction
through de Haas-van Alphen oscillation experiments at $40$mK.
Bianchi \emph{et al.} \cite{Bianchi03} found the maxima of $H_{c2}$
along the [100] direction by measuring the specific heat at $T >
1$K. Weickert \emph{et al.} \cite{Weickert06} measured the electric
resistivity at $100$mK and found the maxima of $H_{c2}$ along the
[100] direction. These measurements seem to agree with each other.
However, Murphy \emph{et al.} \cite{Murphy02} observed the maxima of
$H_{c2}$ along the [110] direction in cantilever magnetometer
measurements performed at $20$mK. At first glance, the observation
of Murphy \emph{et al.} \cite{Murphy02} is in sharp conflict with
other measurements \cite{Settai01, Bianchi03, Weickert06}, and thus
stands as an obstacle in the determination of the precise gap
symmetry of CeCoIn$_{5}$.

Our theoretical analysis suggest that the above experimental results
might be actually well consistent. Note that the measurements of
Murphy \emph{et al.}\cite{Murphy02} are performed at $20$mK, whereas
all the others \cite{Settai01, Bianchi03, Weickert06} are performed
at $T \geq 40$mK. The experimental discrepancy can be naturally
resolved if, as predicted in our analysis, there is a $\pi/4$ shift
in the angular dependence of $H_{c2}$ at certain temperature between
$20$mK and $40$mK. The critical point $T^{*}$ at which $H_{c2}$
shifts by $\pi/4$ can be probed by carefully measuring the angular
dependence of in-plane $H_{c2}$ at a number of different
temperatures falling in the range of $20\mathrm{mK} < T <
40\mathrm{mK}$.

It is also interesting to remark on the behavior of $H_{c2}$ in
CeCu$_{2}$Si$_{2}$. Different from a $d_{x^2 - y^2}$-wave gap
symmetry deduced in most earlier investigations \cite{Stockert08,
Eremin08}, a $d_{xy}$-wave symmetry was proposed by Vieyra \emph{et
al.} \cite{Vieyra11} after comparing model calculations to the
experimental data of $H_{c2}$ measured at $40$mK. This conclusion is
problematic for two reasons. Firstly, as illustrated in our
theoretic analysis, it is not appropriate to fit experimental
results of angle-resolved $H_{c2}$ at some fixed temperature.
Secondly, in the equation of $H_{c2}$ given in Ref.~\cite{Vieyra11},
a rippled cylindrical Fermi surface is adopted. However, no tuning
parameter is adopted in their calculations to characterize how
rippled is the Fermi surface. Our analysis indicate that, in order
to deduce an accurate gap symmetry from the experiments of in-plane
$H_{c2}$, a more reasonable method is to measure the angular
dependence of $H_{c2}$ at a series of temperatures and to see
whether there is a $\pi/4$ shift at certain critical temperature
$T^{\ast}$.

\section{Summary and Discussion\label{Sec:SumDis}}

In summary, we have studied the angular dependence of in-plane upper
critical field $H_{c2}$ in some $d$-wave heavy fermion
superconductors after including both the orbital and Pauli
paramagnetic effects. By solving the equation of $H_{c2}$
systematically, we have showed that whether $H_{c2}$ exhibits its
maxima along the nodal or antinodal direction crucially depends on a
number of tuning parameters in the presence of a strong Pauli
paramagnetic effect. This makes it difficult to entirely fix the
$d$-wave gap symmetry, since a moderate variation of one or some of
the tuning parameter can lead to a $\pi/4$ shift in the angular
dependence of $H_{c2}$. Neglecting Landau level mixing and FFLO
state, we have found a general property that $H_{c2}$ always
exhibits its maxima along the nodal directions at $T < T^*$ and the
antinodal directions at $T^* < T < T_c$, where $T^*$ is a critical
temperature below $T_{c}$, provided that the Pauli paramagnetic
effect is strong enough. When the Landau level mixing is included,
this general property does not qualitatively change. However, the
directions of maxima of $H_{c2}$ take more complex dependence on
temperature if the superconductor has a FFLO ground state.

Our theoretical studies have gone beyond previous works \cite{Won04,
Weickert06, Vieyra11} in several aspects. Firstly, in previous
studies, a rippled cylindrical Fermi surface was employed, but there
is not any effective parameter in the equations of $H_{c2}$ to
characterize the concrete shape of the Fermi surface. In our
analysis, the shape of the rippled cylindrical Fermi surface is
defined by two parameters, namely $\gamma$ and $\lambda$. We have
illustrated via careful calculations that tuning these two
parameters can qualitatively alter the angular dependence of $H_{c2}$.
Secondly, in previous studies, the angular dependence of $H_{c2}$ was
always calculated and then compared with experiments at certain
given temperature. Our analysis have showed that whether the maxima
of $H_{c2}$ are along nodal or antinodal directions is determined by
the values of a number of tuning parameters. Therefore, one should
not identify the correct gap symmetry by measuring the angular
dependence of $H_{c2}$ at a fixed temperature. Instead, measuring
the angular dependence of $H_{c2}$ at a series of temperatures and
examining whether there is a $\pi/4$ shift as the temperature varies
is a better method. Thirdly, our studies provide a candidate
solution to the long-standing experimental discrepancy about the
angular dependence of $H_{c2}$ in CeCoIn$_{5}$.

To gain a more convincing understanding of the angular dependence of
in-plane $H_{c2}$ and its application to realistic experiments of
heavy fermion superconductors, our theoretic analysis may be
improved in several aspects in the future. For instance, an
important assumption used in our analysis is that the
superconducting phase transition is second order, which has also
been used broadly in previous works \cite{Won04, Weickert06,
Vieyra11}. If the phase transition is first order, it would be
difficult to derive an effective equation for $H_{c2}$. In addition,
we have employed in our work an ideal rippled cylindrical Fermi
surface. It would be very interesting to generalize our
consideration to superconductors with a more complicated and more
realistic Fermi surface. We have also ignored the possible influence
of multi-band effects, which was recently found to be important in
several heavy fermion superconductors such as CeCu$_{2}$Si$_{2}$
\cite{Kittaka14, Enayat15, Ikeda15, Tsutsumi15} and CeCoIn$_{5}$
\cite{Allan13, Zhou13}. Finally, our analysis is essentially BCS
mean-field treatment, which neglects correlation effects. The
possible competition and coexistence between superconducting and
antiferromagnetic orders may play some role\cite{Suginishi06,Zwicknagl81} and hence need to be
incorporated in a more refined investigation of $H_{c2}$.

\section*{ACKNOWLEDGEMENTS}

We acknowledge the support by the National Natural Science
Foundation of China under Grants No.11504379, No.11274286,
No.11574285, and No.U1532267.

\appendix

\begin{widetext}

\section{Derivation of equation of $H_{c2}$ in the presence of Landau level mixing
and FFLO state\label{Appendix:Hc2Derivation}}

We will provide the detailed calculations of the equation of
in-plane $H_{c2}$ in the presence of both Landau level mixing and
FFLO state. The linearized gap equation can be written
as\cite{Shimahara97,Shimahara09}
\begin{eqnarray}
-\ln(\frac{T}{T_{c}})\Delta_{\alpha}(\mathbf{R})
&=&\int_{0}^{+\infty}d\eta\frac{\pi T}{\sinh\left(\pi T\eta\right)}
\int_{-\pi}^{\pi}\frac{d\chi}{2\pi}
\int_{0}^{2\pi}\frac{d\varphi}{2\pi}
\left[\gamma_{\alpha}(\hat{\mathbf{k}})\right]^2
\left[\frac{1}{2}-\frac{\exp(ih\eta)}{2}\exp\left(L_{1}\right)
\right] \Delta_{\alpha}(\mathbf{R})\nonumber
\\
&+&\int_{0}^{+\infty}d\eta\frac{\pi T}{\sinh\left(\pi T\eta\right)}
\int_{-\pi}^{\pi}\frac{d\chi}{2\pi}\int_{0}^{2\pi}\frac{d\varphi}{2\pi}
\left[\gamma_{\alpha}(\hat{\mathbf{k}})\right]^2
\left[\frac{1}{2}-\frac{\exp(-ih\eta)}{2}\exp\left(-L_{1}\right)
\right]\Delta_{\alpha}(\mathbf{R}),\label{Eq:AppendixLGqpEq}
\end{eqnarray}
where
\begin{eqnarray}
L_{1}&=&\frac{1}{2}i\eta\sqrt{eH}
\left[\left(v_{a}\sin(\theta-\varphi)+iv_{c}\sin(\chi)\right)a_{+} +
\left(v_{a}\sin(\theta-\varphi) - iv_{c}\sin(\chi)\right)a_{-} +
\sqrt{2}v_{a}\cos(\theta-\varphi)a_{0}\right].
\end{eqnarray}
We assume that $\Delta_{\alpha}(\mathbf{R})$ takes the FFLO state\cite{Weickert06}
\begin{eqnarray}
\Delta_{\alpha}(\mathbf{R}) =
\cos\left(\mathbf{q}\cdot\mathbf{R}\right)
\left[1+C\left(a_{+}\right)^2\right]
\Delta_{0}(\mathbf{R})\label{Eq:AppendixDelta0FFLO}
\end{eqnarray}
with
\begin{eqnarray}
\Delta_{0}(\mathbf{R})&=&
\left(\sqrt{\frac{2eH}{\pi}}\right)^{\frac{1}{2}}
\exp\left[-eH\left(x\sin\theta - y\cos\theta\right)^{2}\right].
\end{eqnarray}
Here, $\mathbf{q}$ is along the direction of the magnetic field,
namely $\mathbf{q} = q\left[\cos\theta\mathbf{e}_{x} +
\sin\theta\mathbf{e}_{y}\right]$. Therefore it is easy to get
$\cos(\mathbf{q}\cdot\mathbf{R}) = \cos\left[q(x\cos\theta
+y\sin\theta )\right]$. One can verify that
\begin{eqnarray}
a_{-}\Delta_{0}=0,
\qquad \left[a_{\pm},\cos(\mathbf{q}\cdot\mathbf{R})\right]=0,
\qquad\left[a_{0},\cos\left(\mathbf{q}\cdot\mathbf{R}\right)\right]
=\sum_{\sigma=\pm1}\frac{\sigma q}{\left(2eH\right)^{\frac{1}{2}}}
\frac{\exp\left[i\sigma q\left(x\cos\theta+y\sin\theta\right)\right]}{2}.
\end{eqnarray}
Substituting Eq.~(\ref{Eq:AppendixDelta0FFLO}) into
Eq.~(\ref{Eq:AppendixLGqpEq}), we obtain
\begin{eqnarray}
&&-\ln\left(\frac{T}{T_{c}}\right)
\cos\left(\mathbf{q}\cdot\mathbf{R}\right) \left[1 +
C\left(a_{+}\right)^2\right] \Delta_{0}(\mathbf{R})\nonumber
\\
&=&\int_{0}^{+\infty}d\eta\frac{\pi T}{\sinh\left(\pi T\eta\right)}
\int_{-\pi}^{\pi}\frac{d\chi}{2\pi}
\int_{0}^{2\pi}\frac{d\varphi}{2\pi}
\left[\gamma_{\alpha}(\hat{\mathbf{k}})\right]^2
\left[\frac{1}{2}-\frac{\exp(ih\eta)}{2}\exp\left(L_{1}\right)
\right]\cos\left(\mathbf{q}\cdot\mathbf{R}\right)
\left[1+C\left(a_{+}\right)^2\right]\Delta_{0}(\mathbf{R})\nonumber
\\
&&+\int_{0}^{+\infty}dt\frac{\pi T}{\sinh\left(\pi T\eta\right)}
\int_{-\pi}^{\pi}\frac{d\chi}{2\pi}\int_{0}^{2\pi}\frac{d\varphi}{2\pi}
\left[\gamma_{\alpha}(\hat{\mathbf{k}})\right]^2
\left[\frac{1}{2}-\frac{\exp(-ih\eta)}{2}\exp\left(-L_{1}\right)
\right]\cos\left(\mathbf{q}\cdot\mathbf{R}\right)
\left[1+C\left(a_{+}\right)^2\right]\nonumber
\\
&&\times\Delta_{0}(\mathbf{R}).\label{Eq:AppenHc2Eq1}
\end{eqnarray}
Exchanging the position of $1 + C\left(a_{+}\right)^2$ and
$\cos\left(\mathbf{q}\cdot\mathbf{R}\right)$ leads us to
\begin{eqnarray}
&&-\ln\left(\frac{T}{T_{c}}\right)\left[1+C\left(a_{+}\right)^2\right]
\cos\left(\mathbf{q}\cdot\mathbf{R}\right)
\Delta_{0}(\mathbf{R})\nonumber
\\
&=&\int_{0}^{+\infty}d\eta\frac{\pi T}{\sinh\left(\pi T\eta\right)}
\int_{-\pi}^{\pi}\frac{d\chi}{2\pi}
\int_{0}^{2\pi}\frac{d\varphi}{2\pi}
\left[\gamma_{\alpha}(\hat{\mathbf{k}})\right]^2
\left\{\frac{1}{2}\left[1+C\left(a_{+}\right)^2\right]
\cos(\mathbf{q}\cdot\mathbf{R})-\frac{\exp(ih\eta)}{2}\right.\nonumber
\\
&&\times\left.L_{3}(\eta)\exp\left(L_{2}\right)L_{4}(\eta)
\right\}\Delta_{0}(\mathbf{R})\nonumber
\\
&&+\int_{0}^{+\infty}d\eta\frac{\pi T}{\sinh\left(\pi T\eta\right)}
\int_{-\pi}^{\pi}\frac{d\chi}{2\pi}\int_{0}^{2\pi}\frac{d\varphi}{2\pi}
\left[\gamma_{\alpha}(\hat{\mathbf{k}})\right]^2
\left\{\frac{1}{2}\left[1+C\left(a_{+}\right)^2\right]
\cos(\mathbf{q}\cdot\mathbf{R})-\frac{\exp(-ih\eta)}{2}\right.\nonumber
\\
&&\left.\times L_{3}(-\eta)
\exp\left(-L_{2}\right)L_{4}(-\eta)\right.\Big\}
\Delta_{0}(\mathbf{R}),\label{Eq:AppenHc2Eq2A}
\end{eqnarray}
where
\begin{eqnarray}
L_{2}(\eta) &=& \frac{i}{2}\eta\sqrt{eH}
\left[\left(v_{a}\sin(\theta-\varphi)+iv_{c}\sin(\chi)\right)a_{+}
+\left(v_{a}\sin(\theta-\varphi)-iv_{c}\sin(\chi)\right)a_{-}\right],
\\
L_{3}(\eta) &=& 1 + C\left[(a_{+})^2 +
i\eta(eH)^{\frac{1}{2}}\left(v_{a}\sin(\theta-\varphi) -
iv_{c}\sin(\chi)\right)a_{+} +
\left(\frac{i}{2}\eta(eH)^{\frac{1}{2}}
\left(v_{a}\sin(\theta-\varphi) -
iv_{c}\sin(\chi)\right)\right)^2\right], \\
L_{4}(\eta)&=&\sum_{\sigma=\pm1}\exp\left[\frac{i}{2}\eta \sigma
v_{a}q\cos(\theta-\varphi)\right]\frac{\exp\left(i\sigma\mathbf{q}
\cdot\mathbf{R}\right)}{2}.
\end{eqnarray}
In the above derivation, we have used the formula
\begin{eqnarray}
[\hat{A},e^{\hat{B}}]=\hat{C}e^{\hat{B}},
\end{eqnarray}
where $\hat{A}$ and $\hat{B}$ do not commute with each other. If
$\hat{C}=[\hat{A},\hat{B}]$, then $\hat{C}$ commutes with $\hat{A}$
and $\hat{B}$, namely $[\hat{A},\hat{C}]=0$ and
$[\hat{B},\hat{C}]=0$. Multiplying $\left(a_{-}\right)^{2}$  on both
sides of the Eq.~(\ref{Eq:AppenHc2Eq1}), and then moving $a_{+}$
leftwards and $a_{-}$ rightwards, we find that
\begin{eqnarray}
&&-\ln\left(\frac{T}{T_{c}}\right)2C \cos\left(\mathbf{q}\cdot
\mathbf{R}\right)\Delta_{0}(\mathbf{R})\nonumber \\
&=&\int_{0}^{+\infty}d\eta\frac{\pi T}{\sinh\left(\pi T\eta\right)}
\int_{-\pi}^{\pi}\frac{d\chi}{2\pi}
\int_{0}^{2\pi}\frac{d\varphi}{2\pi}
\left[\gamma_{\alpha}(\hat{\mathbf{k}})\right]^2
\left[C\cos\left(\mathbf{q}\cdot\mathbf{R}\right) -
\frac{\exp(ih\eta)}{2} L_{5}(\eta)
\exp\left(L_{2}\right)L_{4}(\eta)\right]\Delta_{0}(\mathbf{R})\nonumber
\\
&+&\int_{0}^{+\infty}d\eta\frac{\pi T}{\sinh\left(\pi T\eta\right)}
\int_{-\pi}^{\pi}\frac{d\chi}{2\pi}\int_{0}^{2\pi}\frac{d\varphi}{2\pi}
\left[\gamma_{\alpha}(\hat{\mathbf{k}})\right]^2
\left[C\cos\left(\mathbf{q}\cdot\mathbf{R}\right)-\frac{\exp(-ih\eta)}{2}
L_{5}(-\eta)\exp\left(-L_{2}\right)L_{4}(-\eta)\right]
\Delta_{0}(\mathbf{R}), \nonumber \\
\label{Eq:AppenHc2Eq2B}
\end{eqnarray}
where
\begin{eqnarray}
L_{5}(\eta) &=& 2C + 2C\left[2a_{+}+i\eta(eH)^{1/2}
\left(v_{a}\sin(\theta-\varphi)-iv_{c}\sin(\chi)\right)\right]
\left(\frac{i}{2}\eta(eH)^{1/2}\left(v_{a}
\sin(\theta-\varphi)+iv_{c}\sin(\chi)\right)\right)\nonumber
\\
&&+\left[1+C\left(a_{+}\right)^2+Ci\eta(eH)^{1/2}
\left(v_{a}\sin(\theta-\varphi) -
iv_{c}\sin(\chi)\right)a_{+}\right.\nonumber \\
&&\left.+C\left(\frac{i}{2}\eta(eH)^{1/2}
\left(v_{a}\sin(\theta-\varphi) - iv_{c}
\sin(\chi)\right)\right)^2\right]
\left(\frac{i}{2}\eta(eH)^{1/2}\left(v_{a}
\sin(\theta-\varphi)+iv_{c}\sin(\chi)\right)\right)^2.
\end{eqnarray}
It is necessary to make an average of Eqs.~(\ref{Eq:AppenHc2Eq2A})
and (\ref{Eq:AppenHc2Eq2B}) on the ground state
$\Delta_{0}(\mathbf{R})$. Since $\Delta_{0}(\mathbf{R})$ takes the
form of the eigenfunction of harmonic oscillators, we can use the
formula for harmonic oscillators:
\begin{eqnarray}
\langle e^{\hat{A}} \rangle = e^{\frac{1}{2}\langle
\hat{A}^{2}\rangle},
\end{eqnarray}
and then obtain
\begin{eqnarray}
-\ln\left(\frac{T}{T_{c}}\right) &=&
\int_{0}^{+\infty}d\eta\frac{\pi T}{\sinh\left(\pi T\eta\right)}
\left\{1-\cos(h\eta)\cos\left[\frac{1}{2}\eta
 v_{a}q\cos(\theta-\varphi)\right]\int_{-\pi}^{\pi}\frac{d\chi}{2\pi}
\int_{0}^{2\pi}\frac{d\varphi}{2\pi}
\left[\gamma_{\alpha}(\hat{\mathbf{k}})\right]^2 \right.\nonumber
\\
&&\left.\times\exp\left[-\frac{\eta^2eH}{8}
\left[v_{a}^{2}\sin^{2}(\theta-\varphi)+v_{c}^{2}\sin^{2}(\chi)
\right]\right]\left[1 + \frac{C}{4}\eta^2eH\left(v_{c}^{2}
\sin^{2}(\chi)-v_{a}^{2}\sin^{2}(\theta-\varphi)\right)\right]
\right\}\label{Eq:AppenHc2Eq3A}
\end{eqnarray}
and
\begin{eqnarray}
-\ln\left(\frac{T}{T_{c}}\right)2C
&=&\int_{0}^{+\infty}d\eta\frac{\pi T}{\sinh\left(\pi T\eta\right)}
\left\{2C-\cos(h\eta)\cos\left[\frac{1}{2}\eta v_{a}q\cos(\theta -
\varphi)\right]\int_{-\pi}^{\pi}\frac{d\chi}{2\pi}
\int_{0}^{2\pi}\frac{d\varphi}{2\pi}
\left[\gamma_{\alpha}(\hat{\mathbf{k}})\right]^2\right.\nonumber
\\
&&\times\exp\left[-\frac{\eta^2eH}{8}
\left[v_{a}^{2}\sin^{2}(\theta-\varphi)+v_{c}^{2}\sin^{2}(\chi)
\right]\right]
\left[\frac{1}{4}\eta^2eH\left(v_{c}^{2}\sin^{2}(\chi) -
v_{a}^{2}\sin^{2}(\theta-\varphi)\right)\right.\nonumber
\\
&&\left.\left.+2C\left(1-\frac{1}{2}\eta^2 eH
\left(v_{a}^{2}\sin^{2}(\theta-\varphi) +
v_{c}^{2}\sin^{2}(\chi)\right)
+\frac{1}{32}\eta^4e^{2}H^{2}\left(v_{a}^{2}
\sin^{2}(\theta-\varphi)+v_{c}^{2}
\sin^{2}(\chi)\right)^{2}\right)\right]\right\}.\nonumber \\
\label{Eq:AppenHc2Eq3B}
\end{eqnarray}
Substituting $\left[\gamma_{\alpha}(\hat{\mathbf{k}})\right]^2 = 1 +
\cos(4\varphi)$ into Eqs. (\ref{Eq:AppenHc2Eq3A}) and
(\ref{Eq:AppenHc2Eq3B}) and defining $\pi T \eta=u$, we eventually
find that
\begin{eqnarray}
-\ln(t) &=& \int_{0}^{+\infty}du\frac{1}{\sinh\left(u\right)}
\left\{1 - \cos(hu)\cos\left[pu
\cos(\varphi)\right]\int_{-\pi}^{\pi}\frac{d\chi}{2\pi}
\int_{0}^{2\pi}\frac{d\varphi}{2\pi}
\left[1+\cos(4\varphi)\cos(4\theta)\right]\right.\nonumber
\\
&&\left.\times\exp\left[-\rho u^2
\left(\left(\frac{v_{c}}{v_{a}}\right)^{2}
\sin^{2}(\chi)+\sin^{2}(\varphi) \right)\right]\left[1+2C\rho
u^2\left(\left(\frac{v_{c}}{v_{a}}\right)^{2}
\sin^{2}(\chi)-\sin^{2}(\varphi)\right)\right]\right\}
\end{eqnarray}
and
\begin{eqnarray}
-\ln(t)C &=& \int_{0}^{+\infty}du \frac{1}{\sinh\left(u\right)}
\left\{C-\cos(hu)\cos\left[pu
\cos(\varphi)\right]\int_{-\pi}^{\pi}\frac{d\chi}{2\pi}
\int_{0}^{2\pi}\frac{d\varphi}{2\pi}
\left[1+\cos(4\varphi)\cos(4\theta)\right]\right.\nonumber
\\
&&\times\exp\left[-\rho u^2
\left(\left(\frac{v_{c}}{v_{a}}\right)^{2}
\sin^{2}(\chi)+\sin^{2}(\varphi) \right)\right] \left[\rho
u^2\left(\left(\frac{v_{c}}{v_{a}}\right)^{2}
\sin^{2}(\chi)-\sin^{2}(\varphi)\right)\right.\nonumber \\
&&\left.\left.+C\left(1-4\rho
u^2\left(\left(\frac{v_{c}}{v_{a}}\right)^{2}
\sin^{2}(\chi)+\sin^{2}(\varphi)\right)
+2\rho^2u^4\left(\left(\frac{v_{c}}{v_{a}}\right)^{2}
\sin^{2}(\chi)+\sin^{2}(\varphi)\right)^{2}\right)\right]\right\},
\end{eqnarray}
where
\begin{eqnarray}
t = \frac{T}{T_{c}}, \qquad h = \frac{g\mu_{B}H}{2\pi T},\qquad \rho
= \frac{eHv_{a}^{2}}{8\pi^2 T^2}, \qquad p = \frac{v_{a}q}{2\pi T}.
\end{eqnarray}
\end{widetext}


\begin{thebibliography}{99}

\bibitem{Norman11}
M. R. Norman, Science {\bf 332}, 196 (2011).

\bibitem{Scalapino12}
D. J. Scalapino, Rev. Mod. Phys. {\bf 84}, 1383 (2012).

\bibitem{Gorkov84}
L. P. Gorkov, Pis'ma Zh. Eksp. Theor. Fiz. {\bf 40}, 351 (1984)
[Sov. Phys. JETP Letter {\bf 40}, 1155 (1984)].

\bibitem{Won94}
H. Won and K. Maki, Physica B {\bf 199-200}, 353 (1994).

\bibitem{Takanaka95}
K. Takanaka and K. Kuboya, Phys. Rev. Lett. {\bf 75}, 323 (1995).

\bibitem{Koike96}
Y. Koike, T. Takabayashi, T. Noji, T. Nishizaki, and N. Kobayashi,
Phy. Rev. B {\bf 54}, R776 (1996).

\bibitem{Naito01}
T. Naito, S. Haraguchi, H. Iwasaki, T. Sasaki, T. Nishizaki, K.
Shibata, and N. Kobayashi, Phys. Rev. B {\bf 63}, 172506 (2001).

\bibitem{Won04}
H. Won, K. Maki, S. Haas, N. Oeschler, F. Weickert, P. Gegenwart,
Phys. Rev. B {\bf 69}, 180504(R) (2004).

\bibitem{Weickert06}
F. Weickert, P. Gegenwart, H. Won, D. Parker, and K. Maki, Phys.
Rev. B {\bf 74}, 134511 (2006).

\bibitem{Vieyra11}
H. A. Vieyra, N. Oeschler, S. Seiro, H. S. Jeevan, C. Geibel, D.
Parker, and F. Steglich, Phys. Rev. Lett. {\bf 106}, 207001 (2011).

\bibitem{Murphy13}
J Murphy, M. A. Tanatar, D. Graf, J. S. Brooks, S. L. Bud'ko, P. C.
Canfield, V. G. Kogan, and R. Prozorov, Phys. Rev. B {\bf 87},
094505 (2013).

\bibitem{Mao00}
Z. Q. Mao, Y. Maeno, S. NishiZaki, T. Akima, and T. Ishiguro, Phys.
Rev. Lett. {\bf 84}, 991 (2000).

\bibitem{Zuo15}
H. Zuo, J.-K. Bao, Y. Liu, J. Wang, Z. Jin, Z. Xia, L. Li, Z. Xu, Z.
Zhu, and G.-H. Cao, arXiv:1511.06169v1.

\bibitem{Tsuei00}
C. C. Tsuei and J. R. Kirtley, Rev. Mod. Phys. {\bf 72}, 969 (2000).

\bibitem{Damascelli03}
A. Damascelli, Z. Hussain, and Z.-X. Shen, Rev. Mod. Phys. {\bf 75},
473 (2003).

\bibitem{Lohneysen07}
H. v. L\"{o}hneysen, A. Rosch, M. Vojta, and P. W\"{o}lfle, Rev.
Mod. Phys. {\bf 79}, 1015 (2007).

\bibitem{Stockert12}
O. Stockert, S. Kirchner, F. Steglich, and Q. Si, J. Phys. Soc.
Jpn. {\bf 81}, 011001 (2012).

\bibitem{Matsuda06}
Y. Matsuda, K. Izawa and I. Vekhter, J. Phys.: Condens. Matter {\bf 18}, R705 (2006).

\bibitem{Sarrao07}
J. L. Sarrao and J. D. Thompson, J. Phys. Soc. Jpn. {\bf 76}, 051013 (2007).

\bibitem{Thompson12}
J. D. Thompson and Z. Fisk, J. Phys. Soc. Jpn. {\bf 81}, 011002
(2012).

\bibitem{Steglich79}
F. Steglich, J. Aarts, C. D. Bredl, W. Lieke, D. Meschede, W. Franz,
and H. Sch\"{a}fer, Phys. Rev. Lett. {\bf 43}, 1892 (1979).

\bibitem{Stockert08}
O. Stockert, J. Arndt, A. Schneidewind, H. Schneider, H. S. Jeevan,
C. Geibel, F. Steglich, M. Loewenhaupt, Physica B {\bf 403}, 973
(2008).

\bibitem{Eremin08}
I. Eremin, G. Zwicknagl, P. Thalmeier, and P. Fulde, Phys. Rev.
Lett. {\bf 101}, 187001 (2008).

\bibitem{Kittaka14}
S. Kittaka, Y. Aoki, Y. Shimura, T. Sakakibara, S. Seiro, C. Geibel,
F. Steglich, H. Ikeda, and K. Machida, Phys. Rev. Lett. {\bf 112},
067002 (2014).

\bibitem{Enayat15}
M. Enayat, Z. Sun, A. Maldonado, H. Suderow, S. Seiro, C. Geibel, S.
Wirth, F. Steglich, and P. Wahl, Phys. Rev. B {\bf 93}, 045123 (2016).

\bibitem{Ikeda15}
H. Ikeda, M.-T. Suzuki, and R. Arita, Phys. Rev. Lett. {\bf 114},
147003 (2015).

\bibitem{Pang16}
G. M. Pang, M. Smidman, J. L. Zhang, L. Jiao, Z. F. Weng, E. M. Nica,
Y. Chen, W. B. Jiang, Y. J. Zhang, H. S. Jeevan, P. Gegenwart, F. Steglich,
Q. Si, and H. Q. Yuan, arXiv:1605.04786v1.

\bibitem{Petrovic01}
C. Petrovic, P. G. Pagliuso, M. F. Hundley, R. Movshovich, J. L.
Sarrao, J. D. Thompson, Z. Fisk, and P. Monthoux, J. Phys.: Condens.
Matter {\bf 13}, L337 (2001).

\bibitem{Izawa01}
K. Izawa, H. Yamaguchi, Y. Matsuda, H. Shishido, R. Settai, and
Y. Onuki, Phys. Rev. Lett. {\bf 87}, 057002 (2001).

\bibitem{An10}
K. An, T. Sakakibara, R. Settai, Y. Onuki, M. Hiragi, M. Ichioka,
and K. Machida, Phys. Rev. Lett. {\bf 104}, 037002 (2010).

\bibitem{Park08}
W. K. Park, J. L. Sarrao, J. D. Thompson, and L. H. Greene, Phys.
Rev. Lett. {\bf 100}, 177001 (2008).

\bibitem{Stock08}
C. Stock, C. Broholm, J. Hudis, H. J. Kang, and C. Petrovic, Phys.
Rev. Lett. {\bf 100}, 087001 (2008).

\bibitem{Allan13}
M. P. Allan, F. Massee, D. K. Morr, J. Van Dyke, A. W. Rost, A. P.
Mackenzie, C. Petrovic, and J. C. Davis, Nat. Phys. {\bf 9},
468(2013).

\bibitem{Zhou13}
B. B. Zhou, S. Misra, E. H. da Silva Neto, P. Aynajian, R. E.
Baumbach, J. D. Thompson, E. D. Bauer, and A. Yazdani, Nat. Phys.
{\bf 9}, 474 (2013).

\bibitem{Murphy02}
T. P. Murphy, D. Hall, E. C. Palm, S. W. Tozer, C. Petrovic, Z.
Fisk, R. G. Goodrich, P. G. Pagliuso, J. L. Sarrao, and J. D.
Thompson, Phys, Rev. B {\bf 65}, 100514(R) (2002).

\bibitem{Settai01}
R. Settai, H. Shishido, S. Ikeda, Y. Murakawa, M. Nakashima, D.
Aoki, Y. Haga, H. Harima, and Y. $\mathrm{\bar{O}}$nuki, J. Phys.: Condens Matter
{\bf 13}, L627 (2001).

\bibitem{Bianchi03}
A. Bianchi, R. Movshovich, C. Capan, P. G. Pagliuso, and J. L. Sarrao,
Phys. Rev. Lett. {\bf 91}, 187004 (2003).

\bibitem{Das13}
T. Das, A. B. Vorontsov, I. Vekhter, and M. J. Graf, Phys. Rev. B
{\bf 87}, 174514 (2013).

\bibitem{Bianchi02}
A. Bianchi, R. Movshovich, N. Oeschler, P. Gegenwart, F. Steglich,
J. D. Thompson, P. G. Pagliuso, and J. L. Sarrao, Phys. Rev. Lett.
{\bf 89}, 137002 (2002).

\bibitem{Bianchi08}
A. D. Bianchi, M. Kenzelmann, L. DeBeer-Schmitt, J. S. White, E. M.
Forgan, J. Mesot, M. Zolliker, J. Kohlbrecher, R. Movshovich, E. D.
Bauer, J. L. Sarrao, Z. Fisk, C. Petrovi\'{c}, M. R. Eskildsen, Science
{\bf 319}, 177 (2008).

\bibitem{Vorontsov10}
A. B. Vorontsov and I. Vekhter, Phys. Rev. B {\bf 81}, 094527
(2010).

\bibitem{Fulde64}
P. Fulde and R. A. Ferrell, Phys. Rev. Lett. {\bf 135}, A550 (1964).

\bibitem{Larkin64}
A. I. Larkin and Y. N. Ovchinnikov, Zh. Eksp. Teor. Phys. {\bf 47},
1136 (1964) [Sov. Phys. JETP {\bf 20}, 762 (1965)].

\bibitem{Chubukov}
A. V. Chubukov and L. P. Gor'kov, Phys. Rev. Lett. {\bf 101}, 147004
(2008).

\bibitem{Vorontsov07a}
A. B. Vorontsov and I. Vekhter, Phys. Rev. B {\bf 75}, 224501
(2007).

\bibitem{Thalmeier05}
P. Thalmeier, T. Watanabe, K. Izawa, Y. Matsuda, Phys. Rev. B {\bf
72}, 024539 (2005).

\bibitem{Helfand66}
E. Helfand and N. R. Werthamer, Phys. Rev. {\bf 147}, 288 (1966).

\bibitem{Werthamer66}
N. R. Werthamer, E. Helfand, P. C. Hohenberg, Phys. Rev. {\bf 147},
295 (1966).

\bibitem{Lukyanchuk87}
I. A. Luk'yanchuk and V. P. Mineev, Zh. Eksp. Teor. Phys. \textbf{93}, 2045 (1987) [Sov. Phys. JETP {\bf 66}, 1168
(1987)].

\bibitem{Scharnberg80}
K. Scharnberg and R. A. Klemm, Phys. Rev B {\bf 22}, 5233 (1980).

\bibitem{Shimahara96}
H. Shimahara, S. Matsuo, and K. Nagai, Phys. Rev. B {\bf 53}, 12284
(1996).

\bibitem{Shimahara97}
H. Shimahara and D. Rainer, J. Phys. Soc. Jpn. {\bf 66}, 3591
(1997).

\bibitem{Suginishi06}
Y. Suginishi and H. Shimahara, Phys. Rev. B {\bf 74}, 024518 (2006).

\bibitem{Shimahara09}
H. Shimahara, Phys. Rev. B {\bf 80}, 214512 (2009).

\bibitem{Won96}
H. Won and K. Maki, Phys. Rev. B {\bf 53}, 5927 (1996).

\bibitem{Casalbuoni04}
R. Casalbuoni and G. Nardulli, Rev. Mod. Phys. {\bf 76}, 263 (2004).

\bibitem{Matsuda07}
Y. Matsuda and H. Shimahara, J. Phys. Soc. Jpn. {\bf 76}, 051005 (2007).

\bibitem{Tsutsumi15}
Y. Tsutsumi, K. Machida, and M. Ichioka, Phys. Rev. B {\bf 92},
020502(R) (2015).

\bibitem{Zwicknagl81}
G. Zwicknagl and P. Fulde, Z. Phys. B {\bf 43}, 23 (1981).



\end{thebibliography}
\end{document}